\documentclass[%
superscriptaddress,
amsmath,amssymb,
 aps,prl,
 twocolumn,
 tightenlines,
]{revtex4-1}

\pdfoutput=1

\usepackage{graphicx,url,amssymb,amsmath,longtable,rotating,color,units,wasysym,subfigure,epsfig,hyperref,epstopdf}
\usepackage[dvipsnames]{xcolor}
\usepackage{enumitem}
\usepackage{lineno}

\usepackage{graphicx}
\usepackage{dcolumn}
\usepackage{bm}
\usepackage{url,hyperref}
\usepackage{array,multirow}
\usepackage{color} 
\usepackage{aas_macros} 
\usepackage{ifthen}
\usepackage{etoolbox}

\usepackage{siunitx} 
\definecolor{darkgreen}{rgb}{0.0, 0.5, 0.0}

\newtoggle{fullauthorlist}
\toggletrue{fullauthorlist}

\newtoggle{endauthorlist}
\toggletrue{endauthorlist}

\def\OGW{\Omega_{\rm GW}}
\def\sgwb{stochastic background}
\def\sgwbs{stochastic backgrounds}

\def\HubbleConstant{67.9\ {\rm km s^{-1} Mpc^{-1}}}
\def\fref{{\rm 25\ Hz}}

\def\TimeDomainCutsPercentage{16}
\def\FreqDomainCutsPercentage{15}
\def\LowFreqDomainCutsPercentage{ 4}
\def\HanfordCAL{2.6\%}
\def\LivingstonCAL{3.85\%}
\def\Livetime{99 days}
\def\IsoAlphaZeroPrevUL{1.7 \times 10^{-7}}
\def\IsoAlphaCBCPrevUL{1.3 \times 10^{-7}}
\def\IsoAlphaThreePrevUL{1.7 \times 10^{-8}}

\def\IsoAlphaZeroPtEst{2.2} 
\def\IsoAlphaCBCPtEst{2.0} 
\def\IsoAlphaThreePtEst{3.5} 

\def\IsoAlphaZeroSigma{ 2.2 } 
\def\IsoAlphaCBCSigma{1.6} 
\def\IsoAlphaThreeSigma{2.8} 

\def\PrevIsoAlphaZeroPtEst{4.4} 
\def\PrevIsoAlphaCBCPtEst{3.5} 
\def\PrevIsoAlphaThreePtEst{3.7} 

\def\PrevIsoAlphaZeroSigma{6.0 } 
\def\PrevIsoAlphaCBCSigma{4.4} 
\def\PrevIsoAlphaThreeSigma{6.6} 

\def\improvement{2.8}
\def\IntTimeFactor{2.1}

\def\IsoAlphaZeroFreqs{20-81.9 Hz}
\def\IsoAlphaCBCFreqs{ 20-95.2 Hz}
\def\IsoAlphaThreeFreqs{20-301 Hz}

\def\SNRLOGBFlogUniform{ -0.64}
\def\NGRGRLOGBFlogUniform{ -0.45}

\def\RunOneTwoUniformPriorAlphaZeroLimit{\num{5.99e-08}}

\def\RunOneTwoUniformPriorAlphaTwoThirdsLimit{\num{4.76e-08}}

\def\RunOneTwoUniformPriorAlphaThreeLimit{\num{7.86e-09}}

\def\RunOneLogUniformPriorAlphaZeroLimit{\num{6.37e-08}}

\def\RunOneTwoLogUniformPriorAlphaZeroLimit{\num{3.49e-08}}

\def\RunOneLogUniformPriorAlphaTwoThirdsLimit{\num{5.13e-08}}

\def\RunOneTwoLogUniformPriorAlphaTwoThirdsLimit{\num{2.99e-08}}

\def\RunOneLogUniformPriorAlphaThreeLimit{\num{6.68e-09}}

\def\RunOneTwoLogUniformPriorAlphaThreeLimit{\num{5.05e-09}}

\def\ngrRunOnePolTUniformPriorLimitT{\num{2.52e-07}}

\def\ngrRunOneTwoPolTUniformPriorLimitT{\num{1.13e-07}}

\def\ngrRunOneTwoPolTUniformPriorAlphaT{-1.7}
\def\ngrRunOneTwoPolTUniformPriorAlphaTLowError{5.3}
\def\ngrRunOneTwoPolTUniformPriorAlphaTHighError{4.0}

\def\ngrRunOneTwoPolVUniformPriorLimitV{\num{1.68e-07}}

\def\ngrRunOneTwoPolVUniformPriorAlphaV{-3.6}
\def\ngrRunOneTwoPolVUniformPriorAlphaVLowError{3.6}
\def\ngrRunOneTwoPolVUniformPriorAlphaVHighError{4.3}

\def\ngrRunOneTwoPolSUniformPriorLimitS{\num{5.78e-07}}

\def\ngrRunOneTwoPolSUniformPriorAlphaS{-4.2}
\def\ngrRunOneTwoPolSUniformPriorAlphaSLowError{3.1}
\def\ngrRunOneTwoPolSUniformPriorAlphaSHighError{4.5}

\def\ngrRunOneTwoPolTVUniformPriorLimitT{\num{9.48e-08}}
\def\ngrRunOneTwoPolTVUniformPriorLimitV{\num{1.42e-07}}

\def\ngrRunOneTwoPolTVUniformPriorAlphaT{-1.4}
\def\ngrRunOneTwoPolTVUniformPriorAlphaTLowError{5.4}
\def\ngrRunOneTwoPolTVUniformPriorAlphaTHighError{4.0}
\def\ngrRunOneTwoPolTVUniformPriorAlphaV{-3.4}
\def\ngrRunOneTwoPolTVUniformPriorAlphaVLowError{3.8}
\def\ngrRunOneTwoPolTVUniformPriorAlphaVHighError{4.4}

\def\ngrRunOneTwoPolTSUniformPriorLimitT{\num{9.39e-08}}
\def\ngrRunOneTwoPolTSUniformPriorLimitS{\num{4.81e-07}}

\def\ngrRunOneTwoPolTSUniformPriorAlphaT{-1.2}
\def\ngrRunOneTwoPolTSUniformPriorAlphaTLowError{5.5}
\def\ngrRunOneTwoPolTSUniformPriorAlphaTHighError{3.8}
\def\ngrRunOneTwoPolTSUniformPriorAlphaS{-3.9}
\def\ngrRunOneTwoPolTSUniformPriorAlphaSLowError{3.4}
\def\ngrRunOneTwoPolTSUniformPriorAlphaSHighError{4.6}

\def\ngrRunOneTwoPolVSUniformPriorLimitV{\num{1.44e-07}}
\def\ngrRunOneTwoPolVSUniformPriorLimitS{\num{4.80e-07}}

\def\ngrRunOneTwoPolVSUniformPriorAlphaV{-3.3}
\def\ngrRunOneTwoPolVSUniformPriorAlphaVLowError{3.9}
\def\ngrRunOneTwoPolVSUniformPriorAlphaVHighError{4.3}
\def\ngrRunOneTwoPolVSUniformPriorAlphaS{-3.9}
\def\ngrRunOneTwoPolVSUniformPriorAlphaSLowError{3.4}
\def\ngrRunOneTwoPolVSUniformPriorAlphaSHighError{4.6}

\def\ngrRunOneTwoPolTVSUniformPriorLimitT{\num{8.20e-08}}
\def\ngrRunOneTwoPolTVSUniformPriorLimitV{\num{1.22e-07}}
\def\ngrRunOneTwoPolTVSUniformPriorLimitS{\num{4.16e-07}}

\def\ngrRunOneTwoPolTVSUniformPriorAlphaT{-1.4}
\def\ngrRunOneTwoPolTVSUniformPriorAlphaTLowError{5.4}
\def\ngrRunOneTwoPolTVSUniformPriorAlphaTHighError{3.9}
\def\ngrRunOneTwoPolTVSUniformPriorAlphaV{-3.3}
\def\ngrRunOneTwoPolTVSUniformPriorAlphaVLowError{3.9}
\def\ngrRunOneTwoPolTVSUniformPriorAlphaVHighError{4.4}
\def\ngrRunOneTwoPolTVSUniformPriorAlphaS{-3.7}
\def\ngrRunOneTwoPolTVSUniformPriorAlphaSLowError{3.5}
\def\ngrRunOneTwoPolTVSUniformPriorAlphaSHighError{4.6}

\def\ngrRunOnePolTLogUniformPriorLimitT{\num{5.53e-08}}

\def\ngrRunOneTwoPolTLogUniformPriorLimitT{\num{3.38e-08}}

\def\ngrRunOneTwoPolTLogUniformPriorAlphaT{-0.2}
\def\ngrRunOneTwoPolTLogUniformPriorAlphaTLowError{6.1}
\def\ngrRunOneTwoPolTLogUniformPriorAlphaTHighError{5.3}

\def\ngrRunOneTwoPolVLogUniformPriorLimitV{\num{3.32e-08}}

\def\ngrRunOneTwoPolVLogUniformPriorAlphaV{-0.8}
\def\ngrRunOneTwoPolVLogUniformPriorAlphaVLowError{5.7}
\def\ngrRunOneTwoPolVLogUniformPriorAlphaVHighError{5.4}

\def\ngrRunOneTwoPolSLogUniformPriorLimitS{\num{6.98e-08}}

\def\ngrRunOneTwoPolSLogUniformPriorAlphaS{-0.7}
\def\ngrRunOneTwoPolSLogUniformPriorAlphaSLowError{5.8}
\def\ngrRunOneTwoPolSLogUniformPriorAlphaSHighError{5.8}

\def\ngrRunOneTwoPolTVLogUniformPriorLimitT{\num{3.30e-08}}
\def\ngrRunOneTwoPolTVLogUniformPriorLimitV{\num{3.01e-08}}

\def\ngrRunOneTwoPolTVLogUniformPriorAlphaT{-0.2}
\def\ngrRunOneTwoPolTVLogUniformPriorAlphaTLowError{6.1}
\def\ngrRunOneTwoPolTVLogUniformPriorAlphaTHighError{5.3}
\def\ngrRunOneTwoPolTVLogUniformPriorAlphaV{-0.9}
\def\ngrRunOneTwoPolTVLogUniformPriorAlphaVLowError{5.6}
\def\ngrRunOneTwoPolTVLogUniformPriorAlphaVHighError{5.4}

\def\ngrRunOneTwoPolTSLogUniformPriorLimitT{\num{3.36e-08}}
\def\ngrRunOneTwoPolTSLogUniformPriorLimitS{\num{6.19e-08}}

\def\ngrRunOneTwoPolTSLogUniformPriorAlphaT{-0.2}
\def\ngrRunOneTwoPolTSLogUniformPriorAlphaTLowError{6.2}
\def\ngrRunOneTwoPolTSLogUniformPriorAlphaTHighError{5.3}
\def\ngrRunOneTwoPolTSLogUniformPriorAlphaS{-0.8}
\def\ngrRunOneTwoPolTSLogUniformPriorAlphaSLowError{5.7}
\def\ngrRunOneTwoPolTSLogUniformPriorAlphaSHighError{5.8}

\def\ngrRunOneTwoPolVSLogUniformPriorLimitV{\num{3.17e-08}}
\def\ngrRunOneTwoPolVSLogUniformPriorLimitS{\num{6.39e-08}}

\def\ngrRunOneTwoPolVSLogUniformPriorAlphaV{-0.9}
\def\ngrRunOneTwoPolVSLogUniformPriorAlphaVLowError{5.6}
\def\ngrRunOneTwoPolVSLogUniformPriorAlphaVHighError{5.5}
\def\ngrRunOneTwoPolVSLogUniformPriorAlphaS{-0.8}
\def\ngrRunOneTwoPolVSLogUniformPriorAlphaSLowError{5.7}
\def\ngrRunOneTwoPolVSLogUniformPriorAlphaSHighError{5.9}

\def\ngrRunOneTwoPolTVSLogUniformPriorLimitT{\num{3.20e-08}}
\def\ngrRunOneTwoPolTVSLogUniformPriorLimitV{\num{2.85e-08}}
\def\ngrRunOneTwoPolTVSLogUniformPriorLimitS{\num{6.07e-08}}

\def\ngrRunOneTwoPolTVSLogUniformPriorAlphaT{-0.2}
\def\ngrRunOneTwoPolTVSLogUniformPriorAlphaTLowError{6.1}
\def\ngrRunOneTwoPolTVSLogUniformPriorAlphaTHighError{5.3}
\def\ngrRunOneTwoPolTVSLogUniformPriorAlphaV{-0.8}
\def\ngrRunOneTwoPolTVSLogUniformPriorAlphaVLowError{5.7}
\def\ngrRunOneTwoPolTVSLogUniformPriorAlphaVHighError{5.4}
\def\ngrRunOneTwoPolTVSLogUniformPriorAlphaS{-0.8}
\def\ngrRunOneTwoPolTVSLogUniformPriorAlphaSLowError{5.7}
\def\ngrRunOneTwoPolTVSLogUniformPriorAlphaSHighError{5.8}

\def\lnBayesTvsN{-0.28}
\def\lnBayesVvsN{-0.50}
\def\lnBayesSvsN{-0.42}
\def\lnBayesTVvsN{-0.79}
\def\lnBayesTSvsN{-0.69}
\def\lnBayesVSvsN{-0.92}
\def\lnBayesTVSvsN{-1.20}

\begin{document}
\title{Search for the isotropic stochastic background using data from Advanced LIGO's second observing run}
\date{\today}

\iftoggle{endauthorlist}{
  %
  %
  \let\mymaketitle\maketitle
  \let\myauthor\author
  \let\myaffiliation\affiliation
  \author{The LIGO Scientific Collaboration}
  \author{The Virgo Collaboration}
\noaffiliation
}{
  %
  %
  \iftoggle{fullauthorlist}{
    \input{LSC-Authors-Aug-2018-prd}
  }{
    \author{The LIGO Scientific Collaboration}
    \affiliation{LSC}
    \author{The Virgo Collaboration}
    \affiliation{Virgo}
  }
}

\begin{abstract}
The stochastic gravitational-wave background is a superposition of sources that are either too weak or too numerous to detect individually. In this study we present the results from a cross-correlation analysis on data from Advanced LIGO's second observing run (O2), which we combine with the results of the first observing run (O1). We do not find evidence for a stochastic background, so we place upper limits on the normalized energy density in gravitational waves at the 95\% credible level of $\OGW<\RunOneTwoUniformPriorAlphaZeroLimit$ for a frequency-independent (flat) background and $\OGW< \RunOneTwoUniformPriorAlphaTwoThirdsLimit$ at \fref\ for a background of compact binary coalescences. The upper limit improves over the O1 result by a factor of \improvement. Additionally, we place upper limits on the energy density in an isotropic background of scalar- and vector-polarized gravitational waves, and we discuss the implication of these results for models of compact binaries and cosmic string backgrounds. Finally, we present a conservative estimate of the correlated broadband noise due to the magnetic Schumann resonances in O2, based on magnetometer measurements at both the LIGO Hanford and LIGO Livingston observatories. We find that correlated noise is well below the O2 sensitivity. 
\end{abstract}

\maketitle

\emph{Introduction}---
A superposition of gravitational waves from many astrophysical and cosmological sources creates a stochastic gravitational-wave background (SGWB).  
Sources which may contribute to the \sgwb\ include
compact binary coalescences \citep{2013MNRAS.431..882Z,PhysRevD.84.124037,PhysRevD.85.104024,PhysRevD.84.084004,2011ApJ...739...86Z,rosado_stoch,MarassiEA_2011,ZhuEA_2013}, 
core collapse supernovae \citep{Buonanno:2004tp,Sandick:2006sm,2009MNRAS.398..293M,2010MNRAS.409L.132Z,PhysRevD.72.084001,PhysRevD.73.104024}, 
neutron stars \citep{Ferrari:1998jf,Regimbau:2001kx,PhysRevD.87.063004,2012PhRvD..86j4007R,2011ApJ...729...59Z,Rosado:2012bk,2011MNRAS.411.2549M,2011MNRAS.410.2123H,WuEA_2013,2004MNRAS.351.1237H},
stellar core collapse 
\cite{2015PhRvD..92f3005C,2017PhRvD..95f3015C},
cosmic strings \citep{2005PhRvD..71f3510D,1976JPhA....9.1387K,2002PhLB..536..185S,2007PhRvL..98k1101S,O1_cosmic_strings}, 
primordial black holes \cite{MandicEA_2016,SasakiEA_2016,Wang:2016ana}, 
superradiance of axion clouds around black holes \cite{Brito:2017wnc,Brito:2017zvb,Fan:2017cfw,Tsukada:2018mbp}, 
and gravitational waves produced during inflation \citep{1994PhRvD..50.1157B,1979JETPL..30..682S,2007PhRvL..99v1301E,2012PhRvD..85b3525B,2012PhRvD..85b3534C,Lopez:2013mqa,1997PhRvD..55..435T,2006JCAP...04..010E,axion_inflation}. 
A particularly promising source is the \sgwb\ from compact binary coalescences, especially in light of the detections of one binary neutron star and ten binary black hole mergers  \cite{gw170814,gw170608,gw170104,gw151226,gw150914,o1rates,gw170817,gwtc-1} by the Advanced LIGO Detector, installed in the Laser Interferometer Gravitational-wave Observatory (LIGO) \cite{aLIGO_2015}, and by Advanced Virgo \cite{aVirgo_2015} so far. Measurements of the rate of binary black hole and binary neutron star mergers imply that the \sgwb\ may be large enough to detect with the Advanced LIGO-Virgo detector network \cite{gw150914_stoch,gw170817_stoch}.
The stochastic background is expected to be dominated by compact binaries at redshifts inaccessible to direct searches for gravitational-wave events  \cite{CallisterEA_2016}. Additionally, a detection of the \sgwb\ would enable a model-independent test of general relativity by discerning the polarization of gravitational waves \cite{TestingGR_stoch,stoch_nongr_O1}. Because general relativity predicts only two tensor polarizations for gravitational waves, any detection of alternative polarizations would imply a modification to our current understanding of gravity \cite{Eardley:1973br,Eardley:1974nw,Will:2014kxa}. 
For recent reviews on relevant data analysis methods, see \cite{RomanoCornish,Christensen_2018}.

In this manuscript, we present a search for an isotropic stochastic background using data from Advanced LIGO's second observing run (O2).
As in previous LIGO and Virgo analyses, this search is based on cross-correlating the strain data between pairs of gravitational-wave detectors \cite{stoch_O1,S5StochIso}.
We first review the stochastic search methodology, then describe the data and data quality cuts. As we do not find evidence for the \sgwb, we place upper limits on the possible amplitude of an isotropic \sgwb, as well as limits on the presence of alternative gravitational-wave polarizations.
Upper limits on anisotropic stochastic backgrounds are given in a publication that is a companion to this one \cite{PhysRevD.100.062001}.
We then give updated forecasts of the sensitivities of future stochastic searches and discuss the implications of our current results for the detection of the stochastic background from compact binaries and cosmic strings.
Finally, we present estimates of the correlated noise in the LIGO detectors due to magnetic Schumann resonances \cite{Schumann_theory}, and discuss mitigation strategies that are being pursued for future observing runs.

\emph{Method}---
The isotropic \sgwb\ can be described in terms of the energy density per logarithmic frequency interval
\begin{equation}
\Omega_{\rm GW}(f) = \frac{f}{\rho_c} \frac{{\rm d} \rho_{\rm GW}}{{\rm d} f},
\end{equation}
where ${\rm d}\rho_{\rm GW}$ is the energy density in gravitational waves in the frequency interval from $f$ to $f+{\rm d}f$, and $\rho_c = 3H_0^2 c^2/(8\pi G)$ is the critical energy density required for a spatially flat universe.
Throughout this work we will use the value of the Hubble constant measured by the \emph{Planck} satellite, $H_0 = \HubbleConstant$ \cite{Planck_2015}.

We use the optimal search for a stationary, Gaussian, unpolarized, and isotropic \sgwb, which is the cross-correlation search \cite{christensen92,Allen_Romano_1999,RomanoCornish,Christensen_2018} (however, see \footnote{Strictly speaking, the optimal search would also include the detector auto-correlation in the likelihood, effectively describing subtraction of the noise power spectrum. However, in practice the Advanced LIGO noise spectrum is not known well enough for this approach to be effective.}). For two detectors, we define a cross-correlation statistic $\hat{C}(f)$ in every frequency bin
\begin{equation}
\label{eq:def_cross_corr}
\hat C(f) = \frac{2}{T}  \frac{{\rm Re}[\tilde{s}_1^\star(f) \tilde{s}_2(f)]}{\gamma_T(f) S_0(f)},
\end{equation}
where $\tilde{s}_i(f)$ is the Fourier transform of the strain time series in detector $i=\{1,2\}$, $T$ is the segment duration used to compute the Fourier transform, and $S_{0}(f)$ is the spectral shape for an $\OGW={\rm const}$ background given by
\begin{equation}
S_0(f) = \frac{3 H_0^2}{10\pi^2f^3}.
\end{equation}
The quantity $\gamma_T(f)$ is the normalized overlap reduction function for tensor (T) polarizations \cite{christensen92}, which encodes the geometry of the detectors and acts as a transfer function between strain cross-power and $\OGW(f)$.
Equation \eqref{eq:def_cross_corr} has been normalized so that the expectation value of $\hat C(f)$ is equal to the energy density in each frequency bin
\begin{equation}
\langle \hat C(f) \rangle =\OGW(f).
\end{equation}
In the limit where the gravitational-wave strain amplitude is small compared to instrumental noise, the variance of $\hat C(f)$ is approximately given by
\begin{equation}
\label{eq:def_sigma}
\sigma^2(f) \approx \frac{1}{2T \Delta f}  \frac{P_1(f) P_2(f)}{\gamma_T^2(f) S_0^2(f)},
\end{equation}
where $P_{1,2}(f)$ are the one-sided noise power spectral densities of the two detectors and $\Delta f$ is the frequency resolution, which we take to be $1/32$ Hz. 

An optimal estimator can be constructed for a model of any spectral shape by taking a weighted combination of the cross-correlation statistics across different frequency bins $f_k$
\begin{eqnarray}
\hat \Omega_{\rm ref} &=& \frac{\sum_k w(f_k)^{-1} \hat C(f_k)\sigma^{-2}(f_k)}{\sum_k w(f_k)^{-2}\sigma^{-2}(f_k)}, \nonumber \\
\sigma_\Omega^{-2} &=& \sum_k w(f_k)^{-2} \sigma^{-2}(f_k),
\end{eqnarray}
where the optimal weights for spectral shape $\OGW(f)$ are given by
\begin{equation}
w(f) = \frac{\OGW(f_{\rm ref})}{\OGW(f)}.
\label{eq:optimal-w}
\end{equation}
The broadband estimators are normalized so that $\langle \hat \Omega_{\rm ref}\rangle =\OGW(f_{\rm ref})$.
By appropriate choices of the weights $w(f)$, one may construct an optimal search for \sgwbs\ with arbitrary spectral shapes, or for \sgwbs\ with scalar and vector polarizations.

Many models of the \sgwb\ can be approximated as a power laws \cite{Allen_Romano_1999,StochPE},
\begin{equation}
\Omega_{\rm GW}(f) = \Omega_{\rm ref} \left(\frac{f}{f_{\rm ref}}\right)^\alpha,
\end{equation}
with a spectral index $\alpha$ and an amplitude $\Omega_\mathrm{ref}$ at a reference frequency $f_\mathrm{ref}$. As in the search in Advanced LIGO's first observing run (O1) \cite{stoch_O1}, we will take $f_{\rm ref}=\fref$, which is a convenient choice in the most sensitive part of the frequency band.
While we will seek to generically constrain both $\Omega_\mathrm{ref}$ and $\alpha$ from the data, we will also investigate several specific spectral indices predicted for different gravitational-wave sources.
In the frequency band probed by Advanced LIGO, the \sgwb\ from compact binaries is well approximated by a power law with $\alpha= 2/3$ \cite{Regimbau_2011}.
Slow roll inflation and cosmic string models can be described with $\alpha = 0$ \cite{stoch_cosmo_review_2018}. Finally, following previous analyses \cite{stoch_O1}, we use $\alpha=3$ as an approximate value to stand in for a variety of astrophysical models with positive slopes, such as unresolved supernovae \citep{2009MNRAS.398..293M,2010MNRAS.409L.132Z,PhysRevD.72.084001,PhysRevD.73.104024}.

\emph{Data}---
We analyze data from Advanced LIGO's second observing run, which took place from 16:00:00 UTC on 30 November, 2016 to 22:00:00 UTC on 25 August, 2017. We cross correlate the strain data measured by the two Advanced LIGO detectors, located in Hanford, WA and Livingston, LA in the United States \cite{aLIGO_2015}. Linearly coupled noise has been removed from the strain time series at Hanford and Livingston using Wiener filtering \cite{Noise_sub_O2,Wiener_O2}, see also \cite{Noise_sub_1,Noise_sub_2,Noise_sub_3}. By comparing coherence spectra and narrowband estimators formed with and without Wiener filtering, we additionally verified that this noise subtraction scheme does not introduce correlated artifacts into the Hanford and Livingston data. 

Virgo does not have a significant impact on the sensitivity of the stochastic search in O2, because of the larger detector noise, the fact that less than one month of coincident integration time is available, and because the overlap reduction function is smaller for the Hanford-Virgo and Livingston-Virgo pairs than for Hanford-Livingston. Therefore we do not include Virgo data in the O2 analysis.

The raw strain data are recorded at 16,384 Hz. We first downsample the strain time series to 4096 Hz, and apply a 16th-order high-pass Butterworth filter with knee frequency of 11 Hz to avoid spectral leakage from the noise power spectrum below 20 Hz. Next we apply a Fourier transform to segments with a duration of 192 s, using 50\% overlapping Hann windows, then we coarse-grain six frequency bins to obtain a frequency resolution of 1/32 Hz. As in \cite{stoch_O1}, we observe in the band 20-1726 Hz. The maximum frequency of 1726 Hz is chosen to avoid aliasing effects after downsampling the data.

Next, we apply a series of data quality cuts that remove non-Gaussian features of the data. We remove times when the detectors are known to be unsuitable for science results \footnote{More precisely, we require that both detectors are in observing mode and that no Category 1 vetos are applied \cite{O1_CBC_DQ}.} and times associated with known gravitational-wave events \cite{gwtc-1}. We also remove times where the noise is non-stationary, following the procedure described in the supplement of ~\cite{S5StochIso} (see also \cite{stoch_O1}). These cuts remove \TimeDomainCutsPercentage \% of the coincident time which is in principle suitable for data analysis, leading to a coincident livetime of \Livetime.

In the frequency domain, we remove narrowband coherent lines that are determined to have instrumental or environmental causes, using the methods described in~\cite{CWStochDetchar}. These cuts remove \FreqDomainCutsPercentage \% of the total observing band, but only \LowFreqDomainCutsPercentage\% of the band below 300 Hz, where the isotropic search is most sensitive. The narrow frequency binning of 1/32 Hz was needed to cut out a comb of coherent lines found at integer frequencies.
A list of notch filters corresponding to lines which were removed from the analysis is also available on the public data release page \cite{o2_iso_data_behind_figures}.

\begin{figure}
\includegraphics[width=0.49\textwidth]{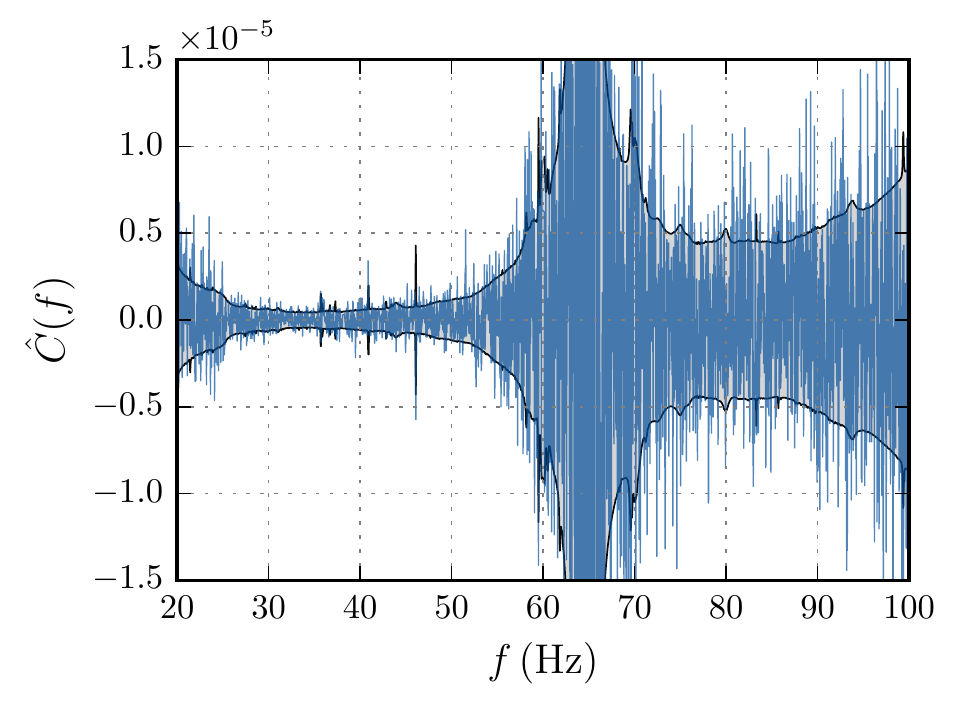}
\caption{The cross-correlation spectrum $\hat C(f)$ measured between  Advanced LIGO's Hanford and Livingston detectors during its second observing run. 
The estimator is normalized so that $\langle \hat C(f) \rangle = \Omega_{\rm GW}(f)$ for tensor-polarized gravitational waves.
The black traces mark the $\pm 1\sigma$ uncertainties on the measured cross-correlations.
Coherent lines that were identified to have an instrumental cause have been removed from the spectrum. The loss in sensitivity visible at approximately 64 Hz is due to a zero in the tensor overlap reduction function $\gamma_T(f)$. }
\label{fig:Y_sig}
\end{figure}

\emph{O2 Results}---
In Figure \ref{fig:Y_sig}, we plot the observed cross-correlation spectrum $\hat C(f)$ and uncertainty $\sigma(f)$ obtained from Advanced LIGO's O2 run. We only plot the spectrum up to 100 Hz to focus on the most sensitive part of the frequency band. These data are also publicly available on the webpage \cite{o2_iso_data_behind_figures}, and can be used to search for \sgwbs\ of any spectral shape.

We perform several tests that the cross-correlation spectrum is consistent with uncorrelated Gaussian noise.
The $\chi^2$ per degree of freedom for the observed spectrum is 0.94.
The loudest individual frequency bin is 51.53 Hz, with a signal-to-noise ratio $C(f)/\sigma(f)$ of 4.2.
With a total of 46,227 (un-notched) frequency bins, there is a 71\% probability that random Gaussian noise would yield an equally loud bin.

In Table ~\ref{tab:ul}, we list the broadband point estimates and 1$\sigma$ uncertainties obtained from the O2 data when assuming power laws with $\alpha=0$, $2/3$, and $3$. Given the uncertainties, uncorrelated Gaussian noise would produce point estimates at least this large with probability 30\%, 22\%, and 21\%, respectively. We conclude there is not sufficient evidence to claim detection of the \sgwb. 

\emph{Upper limits on isotropic stochastic background}---
Since we do not find evidence for the \sgwb, we place upper limits on the amplitude $\Omega_\mathrm{ref}$.
We use the parameter estimation framework described in \cite{StochPE,TestingGR_stoch,stoch_nongr_O1}, applied to the cross-correlation spectrum obtained by combining the results from O1 given in \cite{stoch_O1}, with those from O2 which are described above (please see the Supplementary Material for more details). We present results assuming two priors, one which is uniform in $\Omega_{\rm ref}$ and one which is uniform in $\log{\Omega_{\rm ref}}$.
We additionally marginalize over detector calibration uncertainties \cite{StochCalUncertainty}. In O2 we assume \HanfordCAL\ and \LivingstonCAL\ amplitude uncertainties in Hanford and Livingston, respectively \cite{Cahillane:2017vkb,Viets:2017yvy}. In O1, the calibration uncertainty for Hanford was 4.8\% and for Livingston was 5.4\% \cite{Cahillane:2017vkb}.
Phase calibration uncertainty is negligible.

Figure \ref{fig:ULs} shows the resulting posterior distribution in the $\Omega_{\rm ref}$ vs $\alpha$ plane, along with $68\%$ and $95\%$ credibility contours.
Table \ref{tab:ul2} lists the marginalized 95\% credible upper limit on $\Omega_\mathrm{ref}$ (for both choices of amplitude prior), as well as the amplitude limits obtained when fixing $\alpha=0$, $2/3$, and $3$. 

When adopting a uniform amplitude prior and fixing $\alpha=0$, we obtain an upper limit of $\Omega_{\rm ref}<\RunOneTwoUniformPriorAlphaZeroLimit$, improving the previous O1 result by a factor of \improvement.
The $1\sigma$ error bar is $\IsoAlphaZeroSigma \times 10^{-8}$, a factor of 2.7 times smaller than the equivalent O1 uncertainty.
This factor can be compared with the factor of \IntTimeFactor\ that would be expected based on increased observation time alone, indicating that the search has benefited from improvements in detector noise between O1 and O2. For the compact binary \sgwb\ model of $\alpha=2/3$, we place a limit of $\Omega_{\rm ref}<\RunOneTwoUniformPriorAlphaTwoThirdsLimit$, and for $\alpha=3$, $\Omega_{\rm ref} < \RunOneTwoUniformPriorAlphaThreeLimit$. Finally, when we marginalize over the power law index $\alpha$, we obtain the upper limit $\Omega_{\rm ref}<\ngrRunOneTwoPolTUniformPriorLimitT$. The prior for $\alpha$ is described in the Supplementary Material.

\begin{table}
\setlength{\tabcolsep}{3pt}
\renewcommand{\arraystretch}{1.2}
\begin{tabular}{l|c c | c }
\hline
\hline
$\alpha$ & 
$\hat{\Omega}_{\rm ref}$ (O2)  &
$\hat{\Omega}_{\rm ref}$ (O1)  &
O2 Sensitive band \nonumber \\
\hline
0 &
$(\IsoAlphaZeroPtEst \pm \IsoAlphaZeroSigma) \times 10^{-8}$ & 
$(\PrevIsoAlphaZeroPtEst \pm \PrevIsoAlphaZeroSigma) \times 10^{-8}$ &
\IsoAlphaZeroFreqs \\
2/3 & 
$(\IsoAlphaCBCPtEst \pm \IsoAlphaCBCSigma) \times 10^{-8}$  & 
$(\PrevIsoAlphaCBCPtEst \pm \PrevIsoAlphaCBCSigma) \times 10^{-8}$ & 
\IsoAlphaCBCFreqs  \\
3 & 
$(\IsoAlphaThreePtEst \pm \IsoAlphaThreeSigma) \times 10^{-9}$ & 
$(\PrevIsoAlphaThreePtEst \pm \PrevIsoAlphaThreeSigma) \times 10^{-9}$& 
\IsoAlphaThreeFreqs  \\
\hline
\hline
\end{tabular}
\caption{Point estimates and $1\sigma$ uncertainties for $\Omega_{\rm ref}$ in O2, for different power law models, alongside the same quantities measured in O1 \cite{stoch_O1}. We also show the minimum contiguous frequency band containing 99\% of the sensitivity. For each power law, the maximum of the frequency band is within 5\% of the value found in O1. The value of the Hubble constant used in this paper is different than what was used in the O1 analysis \cite{stoch_O1} ($68 \,\text{km s$^{-1}$ Mpc$^{-1}$}$), which has led to some differences in the numerical values of the point estimates and error bars that we report for O1.}
\label{tab:ul}
\end{table}

\begin{table*}
\setlength{\tabcolsep}{25pt}
\renewcommand{\arraystretch}{1.2}
\begin{tabular}{l|c c|c c}
 &
\multicolumn{2}{c|}{Uniform prior} & 
\multicolumn{2}{c}{Log-uniform prior} \\
\hline
\hline
$\alpha$ &
O1+O2 &
O1 &
O1+O2 &
O1 \\
\hline
0 &
$\RunOneTwoUniformPriorAlphaZeroLimit$ &
$\IsoAlphaZeroPrevUL$& 
$\RunOneTwoLogUniformPriorAlphaZeroLimit$ &
$\RunOneLogUniformPriorAlphaZeroLimit$\\
2/3 &
$\RunOneTwoUniformPriorAlphaTwoThirdsLimit$ &
$\IsoAlphaCBCPrevUL$ & 
$\RunOneTwoLogUniformPriorAlphaTwoThirdsLimit$ & 
$\RunOneLogUniformPriorAlphaTwoThirdsLimit$ \\
3 &
$\RunOneTwoUniformPriorAlphaThreeLimit$ &
$\IsoAlphaThreePrevUL$ & 
$\RunOneTwoLogUniformPriorAlphaThreeLimit$ & 
$\RunOneLogUniformPriorAlphaThreeLimit$ \\
Marg. &
$\ngrRunOneTwoPolTUniformPriorLimitT$ &
$\ngrRunOnePolTUniformPriorLimitT$ & 
$\ngrRunOneTwoPolTLogUniformPriorLimitT$ & 
$\ngrRunOnePolTLogUniformPriorLimitT$ \\
\hline
\hline
\end{tabular}
\caption{95\% credible upper limits on $\Omega_{\rm ref}$ for different power law models (fixed $\alpha$), as well as marginalizing over $\alpha$, for combined O1 and O2 data (current limits) and for O1 data (previous limits) \cite{stoch_O1}. We show results for two priors, one which is uniform in $\Omega_{\rm ref}$, and one which is uniform in the logarithm of $\Omega_{\rm ref}$.}
\label{tab:ul2}
\end{table*}

\begin{figure}
\includegraphics[width=0.49\textwidth]{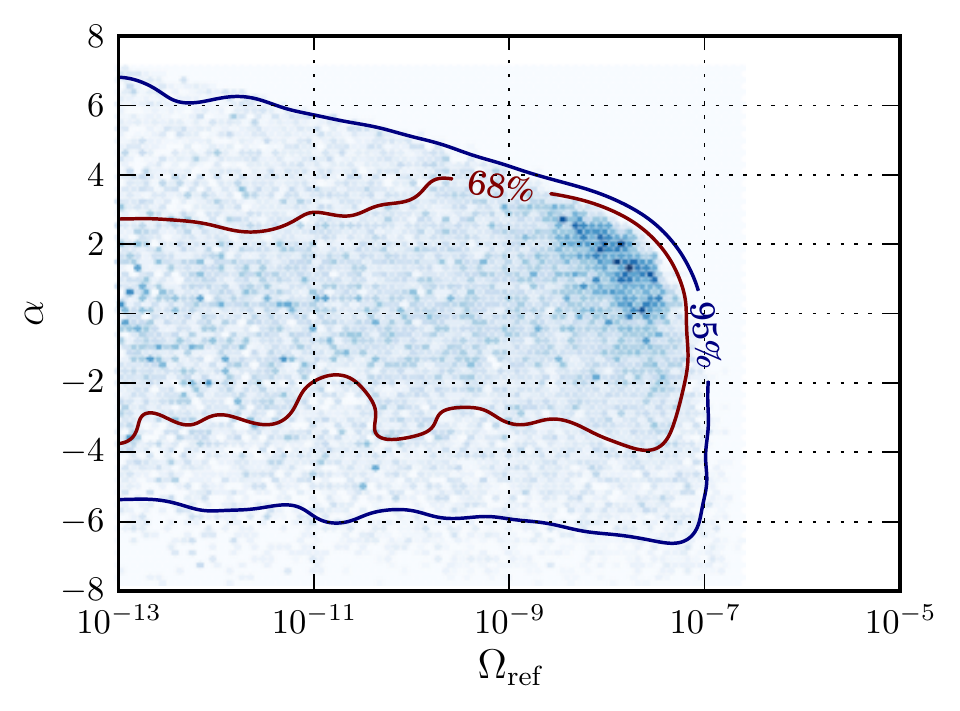}
\caption{Posterior distribution for the amplitude $\Omega_{\rm ref}$ and slope $\alpha$ of the \sgwb, using a prior which is uniform in the logarithm of $\Omega_{\rm ref}$, along with contours with 68\% and 95\% confidence level, using combined O1 and O2 data. There is a small region of increased posterior probability centered around $\log \Omega_{\rm ref}=-8$ and $\alpha=2$. This is not statistically significant, and similar-size bumps have appeared in simulations of Gaussian noise. An analogous plot with a prior uniform in $\Omega_{\rm ref}$ can be found in the Supplementary Material.}
\label{fig:ULs}
\end{figure}

\begin{table}
\setlength{\tabcolsep}{10pt}
\renewcommand{\arraystretch}{1.2}
\begin{tabular}{l|c c}
\hline
\hline
Polarization &
Uniform prior &
Log-uniform prior  \\
\hline
Tensor & $\ngrRunOneTwoPolTVSUniformPriorLimitT$ & $\ngrRunOneTwoPolTVSLogUniformPriorLimitT$ \nonumber \\
Vector & $\ngrRunOneTwoPolTVSUniformPriorLimitV$ & $\ngrRunOneTwoPolTVSLogUniformPriorLimitV$ \nonumber \\
Scalar & $\ngrRunOneTwoPolTVSUniformPriorLimitS$ & $\ngrRunOneTwoPolTVSLogUniformPriorLimitS$ \nonumber \\
\hline
\hline
\end{tabular}
\caption{Upper limits on different polarizations. To obtain the upper limits, we assume a log uniform and a uniform prior on the amplitude $\Omega_{\rm ref}$ for each polarization, using combined O1 and O2 data. We assume the presence of a tensor, vector, and scalar backgrounds, then marginalize over the spectral indices and two amplitudes for the three different polarization modes, as described in the main text.}
\label{tab:nongrUL}
\end{table}

\emph{Implications for compact binary background}---
In Figure \ref{fig:cbc} we show the prediction of the astrophysical \sgwb\ from binary black holes (BBH) and binary neutron stars (BNS), along with its statistical uncertainty due to Poisson uncertainties in the local binary merger rate.
We plot the upper limit allowed from adding the background from neutron star-black hole (NSBH) binaries as a dotted line.
We use the same binary formation and evolution scenario to compute the \sgwb\ from BBH and BNS as in \cite{gw170817_stoch}, but we have updated the mass distributions and rates to be consistent with the most recent results given in \cite{gwtc-1,RatesAndPop_O2}. For NSBHs, we use the same evolution with redshift as BNSs. As in \cite{gw170817}, for BBH we include inspiral, merger and ringdown contributions computed in \cite{AjithEA_2008}, while for NSBH and BNS we use only the inspiral part of the waveform.
For the BBH mass distribution, we assume a power law in the primary mass $p(m_1)\propto m_1^{-2.3}$ with the secondary mass drawn from a uniform distribution, subject to the constraints $5 M_\odot \leq m_2 \leq m_1 \leq 50 M_\odot$. In Ref.~\cite{gwtc-1}, rate estimates were computed by two pipelines, PyCBC \cite{pycbc-1} and GstLAL \cite{gstlal-1}. We use the merger rate measured by GstLAL, $R_{\rm local}=56^{+44}_{-27}{\rm Gpc^{-3} yr^{-1}}$ \cite{gwtc-1}, because it gives a more conservative (smaller) rate estimate. Using the methods described in \cite{gw170817_stoch}, the inferred amplitude of the \sgwb\ is $\Omega_{\rm BBH}(25\ {\rm Hz})=5.3^{+4.2}_{-2.5}\times 10^{-10}$.

For the BNS mass distribution, following the analysis in \cite{gwtc-1}, we take each component mass to be drawn from a Gaussian distribution with a mean of $1.33 M_\odot$ and a standard deviation of $0.09 M_\odot$. We use the GstLAL rate of $R_{\rm local}=920^{+2220}_{-790}{\rm Gpc^{-3} yr^{-1}}$ \cite{gwtc-1}. From these inputs, we predict $\Omega_{\rm BNS}(25\ \rm Hz)=3.6^{+8.4}_{-3.1}\times 10^{-10}$. 
Combining the BBH and BNS results yields a prediction for the total SGWB of $\Omega_{\rm BBH+BNS}(25\ \rm Hz)=8.9^{+12.6}_{-5.6}\times 10^{-10}$. This value is about a factor of 2 smaller the one in \cite{gw170817_stoch}, due in part to the decrease in the rate measured after analyzing O1 and O2 data with the best available sensitivity and data analysis techniques.

For NSBH we assume a delta function mass distribution, where the neutron star has a mass of 1.4 $M_\odot$ and the black hole has a mass of 10 $M_\odot$, and we take the upper limit on the rate from GstLAL \cite{gwtc-1}.
The upper limit from NSBH is $\Omega_{\rm NSBH}(25\ {\rm Hz})=9.1\times 10^{-10}$. We show the sum of the upper limit of $\Omega_{\rm NSBH}(f)$, with the 90\% upper limit on $\Omega_{\rm BBH+BNS}(f)$, as a dotted line in Figure~\ref{fig:cbc}.

We also show the power-law-integrated curves (PI curves) \cite{locus} of the O1 and O2 isotropic background searches.
A power-law \sgwb\ that is tangent to a PI curve is detectable with ${\rm SNR} = 2$ by the given search.
We additionally show a projected PI curve based on operating Advanced LIGO and Advanced Virgo at design sensitivity for 2 years, with 50\% network duty cycle. 
By design sensitivity, we refer to a noise curve which is determined by fundamental noise sources.
We use the Advanced LIGO design sensitivity projection given in \cite{new_design}, which incorporates improved measurements of coating thermal noise relative to the one assumed in \cite{gw150914_stoch}. This updated curve introduces additional broadband noise at low frequencies relative to previous estimates. As a result, the updated design-sensitivity PI curve is less sensitive than the one shown in \cite{gw150914_stoch}.

\begin{figure}
\includegraphics[width=0.49\textwidth]{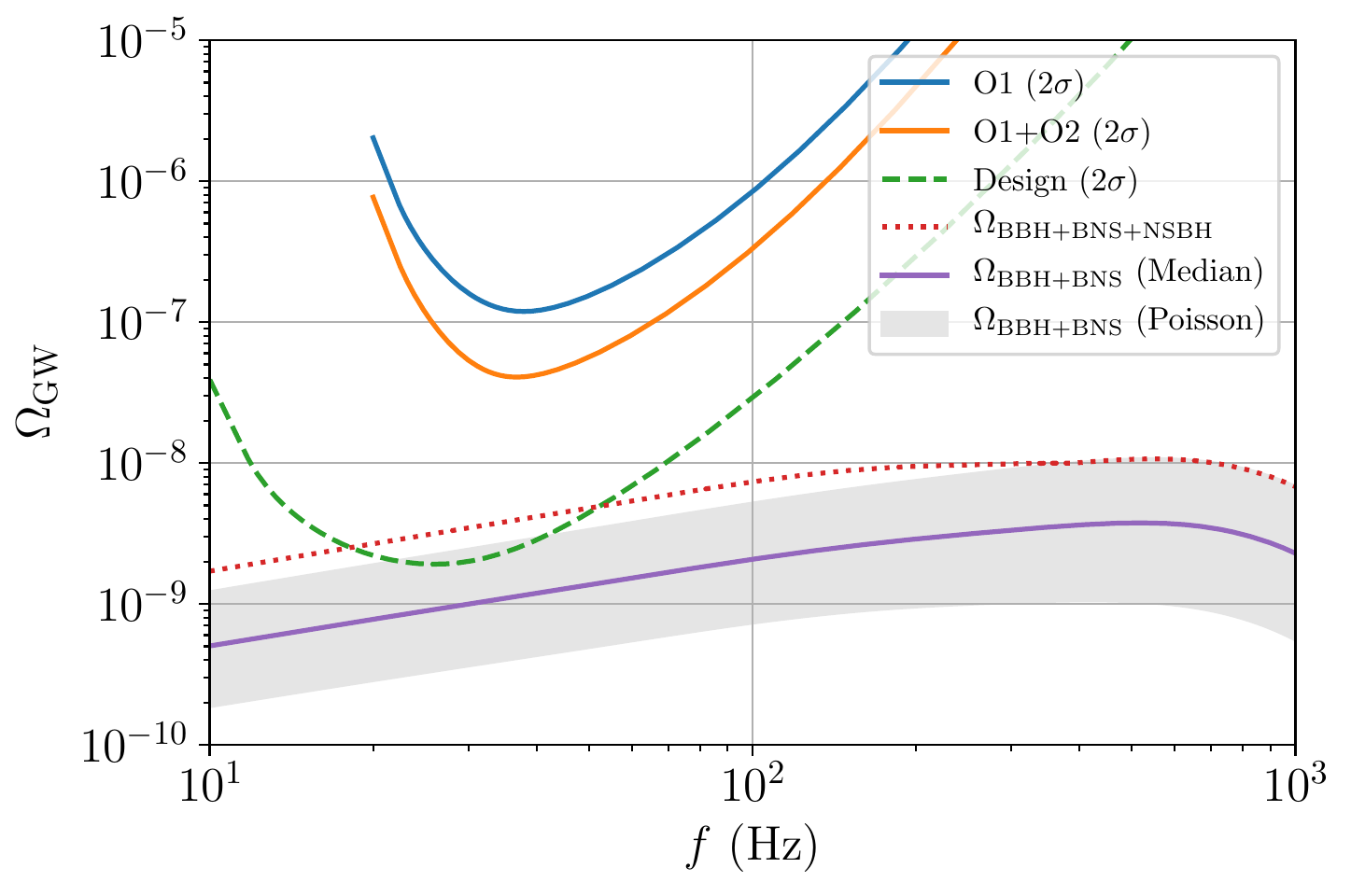}
\caption{Sensitivity curves for O1, combined O1+O2, and design sensitivity. A power law \sgwb\ which lies tangent to one of these curves is detectable with 2$\sigma$ significance. We have used the Advanced LIGO design sensitivity given in \cite{new_design}, which incorporates improved measurements of coating thermal noise. Design sensitivity assumes that the LIGO noise curve is determined by fundamental noise sources only. The purple line is the median total \sgwb, combining BBH and BNS, using the model described in \cite{gw170817_stoch} with updated mass distributions and rates from \cite{gwtc-1,RatesAndPop_O2}, and the gray box is the Poisson error region. The dotted gray line is the sum of the upper limit for the BBH+BNS backgrounds with the upper limit on the NSBH background.}
\label{fig:cbc}
\end{figure}

\emph{Implications for cosmic string models} ---
Cosmic strings~\cite{Kibble:1976sj,Vilenkin:2000jqa} are linear topological defects which are expected to be generically produced within the context of Grand Unified Theories~\cite{Jeannerot:2003qv}.
The dynamics of a cosmic string network is driven by the formation of loops and the emission of gravitational waves~\cite{Vachaspati:1984gt,Sakellariadou:1990ne}. One may therefore use the \sgwb\ in order to constrain the parameters of a cosmic string network.

We will focus on Nambu-Goto strings \cite{NG_Nambu,NG_Goto}, for which the string thickness is zero and the intercommutation probability equals unity.
Gravitational waves will allow us to constrain the string tension $G\mu/c^2$, where $\mu$ denotes the mass per unit length. This dimensionless parameter is the single quantity that characterizes a Nambu-Goto string network.

We will consider two analytic models of cosmic string loop distributions~\cite{Blanco-Pillado:2013qja,Lorenz:2010sm}. The former~\cite{Blanco-Pillado:2013qja} gives the distribution of string
loops of given size at fixed time, under the assumption that the momentum dependence of the loop
production function is weak. The latter~\cite{Lorenz:2010sm} is based on a different numerical simulation~\cite{Ringeval:2005kr}, and gives the distribution of non-self intersecting loops at a given time~\footnote{These models are dubbed model $M=2$ and model $M=3$ in~\cite{O1_cosmic_strings}. 
We do not discuss model $M=1$ of ~\cite{O1_cosmic_strings}, which assumes that all loops are formed with the same relative size, since such a hypothesis is not supported by any numerical simulation of Nambu-Goto string networks.}.

The corresponding limits found by combining O1 and O2 data are $G\mu/c^2\leq 1.1 \times 10^{-6}$ for the model of \cite{Blanco-Pillado:2013qja} and $G\mu/c^2\leq 2.1 \times 10^{-14}$ for the model of \cite{Lorenz:2010sm}. The Advanced LIGO constraints are stronger for the model of \cite{Lorenz:2010sm} because the predicted spectrum is larger at 100 Hz for that model.
This can be compared with the pulsar timing limits, $G\mu/c^2 \leq 1.6 \times 10^{-11}$ and $G\mu/c^2 \leq 6.2 \times 10^{-12}$, respectively \cite{Lasky:2015lej}.

\emph{Test of General Relativity}---
Alternative theories of gravity generically predict the presence of vector or scalar gravitational-wave polarizations in addition to the standard tensor polarizations allowed in general relativity.
Detection of the \sgwb\ would allow for direct measurement of its polarization content, enabling new tests of general relativity \cite{TestingGR_stoch,stoch_nongr_O1}.

When allowing for the presence of alternative gravitational-wave polarizations, the expectation value of the cross-correlation statistic becomes
\begin{equation}
\langle \hat C(f) \rangle = \sum_{A} \beta_{A}(f) \Omega_{\rm GW}^A(f) = \sum_A \beta_A(f) \Omega_{\rm ref}^A \left(\frac{f}{f_{\rm ref}}\right)^{\alpha_A},
\label{eq:nongrY}
\end{equation}
where $\beta_A = \gamma_A(f)/\gamma_T(f)$, and $A$ labels the polarization, $A=\{T,V,S\}$. The functions  $\gamma_T(f)$, $\gamma_V(f)$, and $\gamma_S(f)$ are the overlap reduction functions for tensor, vector, and scalar polarizations \cite{TestingGR_stoch}.
Because these overlap reduction functions are distinct, the spectral shape of $\hat C(f)$ enables us to infer the polarization content of the \sgwb.
While we use the notation $\Omega^A_{\rm GW}(f)$ in analogy with the general relativity (GR) case, in a general modification of gravity, the quantities $\Omega^T_\mathrm{GW}(f)$, $\Omega^V_\mathrm{GW}(f)$, and $\Omega^S_\mathrm{GW}(f)$ are best understood as a measurement of the two-point correlation statistics of different components of the \sgwb\ rather than energy densities \cite{Isi_Stein_NonGR_Energy}.

Following Refs.~\cite{TestingGR_stoch,stoch_nongr_O1}, we compute two Bayesian odds: odds $\mathcal{O}^\textsc{s}_\textsc{n}$ for the presence of a stochastic signal of any polarization(s) versus Gaussian noise, and odds $\mathcal{O}^\textsc{ngr}_\textsc{gr}$ between a hypothesis allowing for vector and scalar modes and a hypothesis restricting to standard tensor polarizations.
Using the combined O1 and O2 measurements, we find $\log\mathcal{O}^\textsc{s}_\textsc{n} = \SNRLOGBFlogUniform$ and $\log\mathcal{O}^\textsc{ngr}_\textsc{gr} = \NGRGRLOGBFlogUniform$, consistent with Gaussian noise.
Given the non-detection of any generic stochastic background, we use Eq.~\eqref{eq:nongrY} to place improved upper limits on the tensor, vector, and scalar background amplitudes, after marginalizing over all three spectral indices, using the priors described in the Supplementary Material.
These limits are shown in Table \ref{tab:nongrUL}, again for both choices of amplitude prior.

\emph{Estimate of correlated magnetic noise}---
Coherent noise between gravitational-wave interferometers may be introduced by terrestrial sources such as Schumann resonances, which are global electromagnetic modes of the cavity formed by the Earth's surface and ionosphere \cite{Schumann_theory}. These fields have very long coherence lengths \cite{Schumann_4} and can magnetically couple to the gravitational-wave channel and lead to broadband noise that is coherent between different gravitational-wave detectors. As the detectors become more sensitive, eventually this source of correlated noise may become visible to the cross-correlation search, and, if not treated carefully, will bias the analysis by appearing as an apparent \sgwb. Unlike the lines and combs discussed in \cite{CWStochDetchar}, we cannot simply remove affected frequency bins from the analysis because Schumann noise is broadband.

Here, we estimate the level of correlated electromagnetic  noise (from Schumann resonances or other sources) in O2 following \cite{Schumann_1,Schumann_2,stoch_O1}. We first measure the cross-power spectral density $M_{12}(f)$ between two Bartington Model MAG-03MC magnetometers \cite{bartington} installed at Hanford and Livingston. We then estimate the transfer function $T_{i}(f)$ ($i=\{1,2\}$) between the magnetometer channel and the gravitational-wave channel at each site, as described in \cite{mag_TF_measurements}.  Finally, we combine these results to produce an estimate for the amount of correlated magnetic noise, which we express in terms of an effective gravitational-wave energy density $\Omega_{\rm mag}(f)$
\begin{equation}
\Omega_{\rm mag}(f) = \frac{ |T_1(f)| |T_2(f)| {\rm Re}[M_{12}(f)]}{\gamma_T(f) S_0(f)} .
\end{equation}
We show $\Omega_{\rm mag}(f)$ in Fig.~\ref{fig:schumann}, alongside the measured O1+O2 PI curve and the projected design-sensitivity PI curve.
The trend for the magnetic noise lies significantly below the O1+O2 PI curve, indicating that correlated magnetic noise is more than an order of magnitude below the sensitivity curve in O2, although it may be an issue for future runs. Experimental improvements can mitigate this risk by further reducing the coupling of correlated noise.
From O1 to O2, for instance, the magnetic coupling was reduced by approximately an order of magnitude, as indicated by the dotted and dot-dashed curves in Fig.~\ref{fig:schumann}. Additionally, work is ongoing to develop Wiener filtering to subtract Schumann noise  \cite{Schumann_2,Schumann_3,Schumann_4}, and to develop a parameter estimation framework to measure or place upper limits on the level of magnetic contamination \cite{meyers_thesis}. This work will take advantage of low noise LEMI-120 magnetometers \cite{lemi} that were recently installed at both LIGO sites, as described in the Supplementary Material.

\begin{figure}
\includegraphics[width=0.49\textwidth]{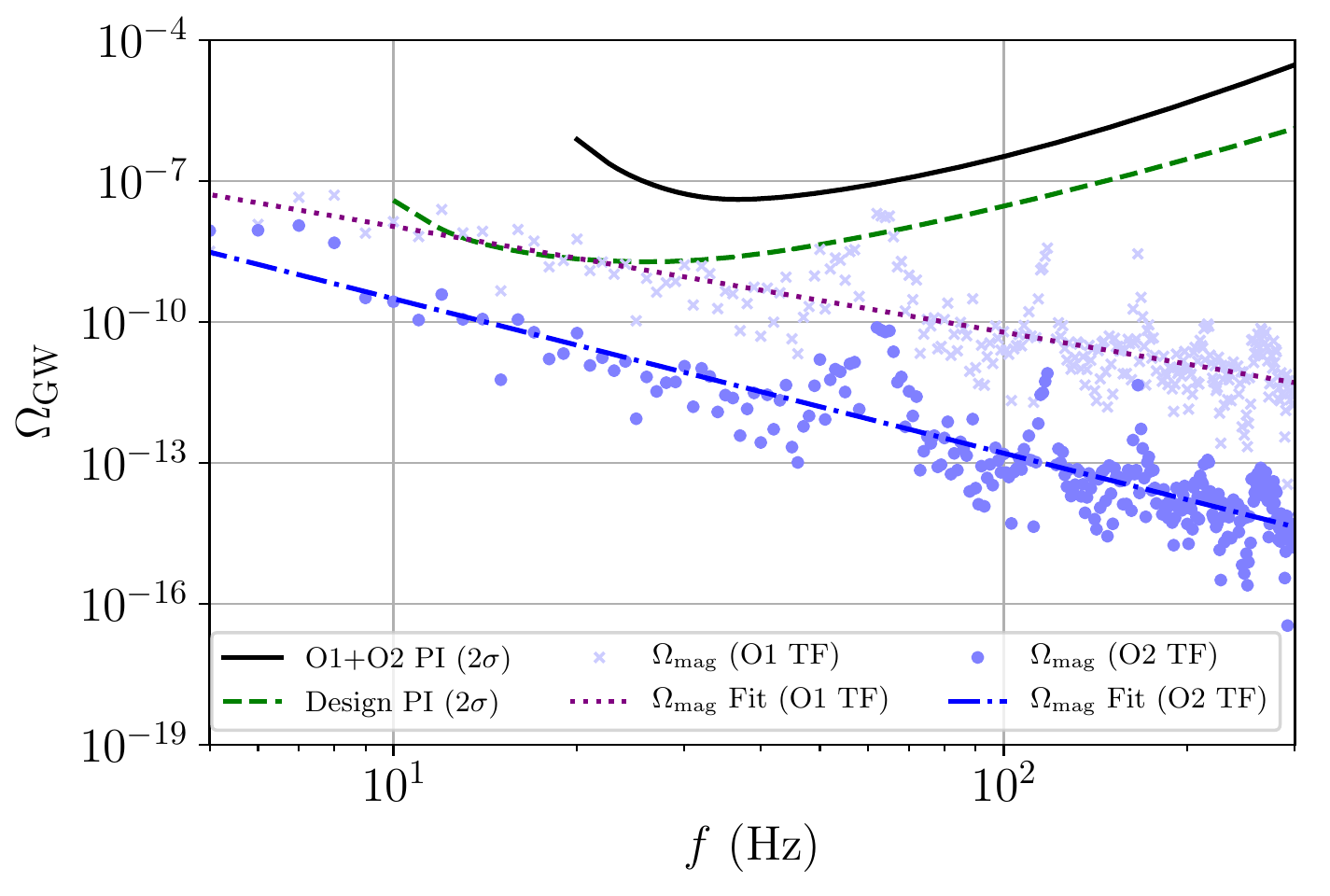}
\caption{Conservative estimate of correlated magnetic noise. We assume a conservative transfer function (TF) based on measurements as described in the text. The first Schumann resonance at 8 Hz is visible, higher harmonics are below the noise floor. There is a zero of the overlap function at 64 Hz which leads to an apparent feature in $\Omega_{\rm mag}$. Power line harmonics have been removed, as in the cross-correlation analysis. The two trend lines show power law fits to the magnetometer spectra, scaled by the O1 (purple dotted) and end-of-O2 (blue dot-dashed) transfer functions. This demonstrates the effect of reducing the magnetic coupling in O2. The trend for the noise budget lies well below the solid black O2 PI curve, which indicates that correlated magnetic noise is negligible in O2. However magnetic contamination may be an issue in future observing runs.}
\label{fig:schumann}
\end{figure}

\emph{Conclusions}---
We have presented the results of a cross-correlation search for the isotropic stochastic background using data from Advanced LIGO's first and second observing runs. While we did not find evidence for the \sgwb, we obtain the most sensitive upper limits to date in the $\sim$20-100 Hz frequency band. We have also placed improved upper limits on the existence of a \sgwb\ from vector and scalar-polarized gravitational waves.

While the upper limits on the SGWB presented in this work are the strongest 
direct limits in the frequency band of current ground-based gravitational-wave 
detectors, other observations place stronger constraints in other frequency 
bands. The NANOGrav collaboration has reported the 95\% upper limit of 
$\Omega_{\rm GW} < 7.4\times 10^{-10}$ at a frequency of $1\ {\rm yr}^{-1}$ after marginalizing over uncertainty in the solar system ephemeris \cite{Arzoumanian:2018saf}. Combining data from the \emph{Planck} satellite and the BICEP2/Keck array constrain the tensor-to-scalar ratio from the CMB to be $r<0.064$ at 95\% confidence at comoving scales of $k=0.002\ {\rm Mpc^{-1}}$, corresponding to a gravitational wave frequency of $f_{0.002}=(2\pi)^{-1} ck=3.1\times 10^{-18} {\rm Hz}$ \cite{Akrami:2018odb}, assuming the single field slow roll consistency condition. Using Equation 4 of \cite{Lasky:2015lej}, this can be converted into the constraint $\Omega_{\rm GW}(f)\leq 3.2 \times 10^{-16} \times (f/f_{\rm 0.05})^{-r/8}\left[16/9+f_{\rm eq}^{2}/(2f^2)\right]$, where $f_{\rm eq}$ is the frequency of a gravitational wave of which the wavelength was the size of the Universe at matter-radiation equality, and $f_{\rm 0.05}$ is the pivot scale. Combining constraints at different frequency ranges can probe models which span many orders of magnitude 
in frequency \cite{Lasky:2015lej,Akrami:2018odb}.

While we have targeted an isotropic, stationary, and Gaussian background, other search techniques can probe backgrounds that violate one or more of these assumptions.
Upper limits on an anisotropic gravitational-wave background from O1 were presented in \cite{stoch_dir_O1}. Furthermore, non-Gaussian searches targeting the compact binary \sgwb\ are currently being developed \cite{popcorn1,popcorn2,popcorn3,tbs_methods}. A successful detection of the \sgwb\ by any of these approaches would offer a new probe of the gravitational-wave sky.

\emph{Acknowledgments} --- 
The authors gratefully acknowledge the support of the United States
National Science Foundation (NSF) for the construction and operation of the
LIGO Laboratory and Advanced LIGO as well as the Science and Technology Facilities Council (STFC) of the
United Kingdom, the Max-Planck-Society (MPS), and the State of
Niedersachsen/Germany for support of the construction of Advanced LIGO 
and construction and operation of the GEO600 detector. 
Additional support for Advanced LIGO was provided by the Australian Research Council.
The authors gratefully acknowledge the Italian Istituto Nazionale di Fisica Nucleare (INFN),  
the French Centre National de la Recherche Scientifique (CNRS) and
the Foundation for Fundamental Research on Matter supported by the Netherlands Organisation for Scientific Research, 
for the construction and operation of the Virgo detector
and the creation and support  of the EGO consortium. 
The authors also gratefully acknowledge research support from these agencies as well as by 
the Council of Scientific and Industrial Research of India, 
the Department of Science and Technology, India,
the Science \& Engineering Research Board (SERB), India,
the Ministry of Human Resource Development, India,
the Spanish  Agencia Estatal de Investigaci\'on,
the Vicepresid\`encia i Conselleria d'Innovaci\'o, Recerca i Turisme and the Conselleria d'Educaci\'o i Universitat del Govern de les Illes Balears,
the Conselleria d'Educaci\'o, Investigaci\'o, Cultura i Esport de la Generalitat Valenciana,
the National Science Centre of Poland,
the Swiss National Science Foundation (SNSF),
the Russian Foundation for Basic Research, 
the Russian Science Foundation,
the European Commission,
the European Regional Development Funds (ERDF),
the Royal Society, 
the Scottish Funding Council, 
the Scottish Universities Physics Alliance, 
the Hungarian Scientific Research Fund (OTKA),
the Lyon Institute of Origins (LIO),
the Paris \^{I}le-de-France Region, 
the National Research, Development and Innovation Office Hungary (NKFIH), 
the National Research Foundation of Korea,
Industry Canada and the Province of Ontario through the Ministry of Economic Development and Innovation, 
the Natural Science and Engineering Research Council Canada,
the Canadian Institute for Advanced Research,
the Brazilian Ministry of Science, Technology, Innovations, and Communications,
the International Center for Theoretical Physics South American Institute for Fundamental Research (ICTP-SAIFR), 
the Research Grants Council of Hong Kong,
the National Natural Science Foundation of China (NSFC),
the Leverhulme Trust, 
the Research Corporation, 
the Ministry of Science and Technology (MOST), Taiwan
and
the Kavli Foundation.
The authors gratefully acknowledge the support of the NSF, STFC, MPS, INFN, CNRS and the
State of Niedersachsen/Germany for provision of computational resources.

This article has been assigned the document number LIGO-P1800258.

\bibliographystyle{apsrev4-1} 
\bibliography{refs}

\iftoggle{endauthorlist}{
  \let\author\myauthor
  \let\affiliation\myaffiliation
  \let\maketitle\mymaketitle
  \title{The LIGO Scientific Collaboration and Virgo Collaboration}
  \pacs{}
  \author{B.~P.~Abbott}
\affiliation{LIGO, California Institute of Technology, Pasadena, CA 91125, USA}
\author{R.~Abbott}
\affiliation{LIGO, California Institute of Technology, Pasadena, CA 91125, USA}
\author{T.~D.~Abbott}
\affiliation{Louisiana State University, Baton Rouge, LA 70803, USA}
\author{S.~Abraham}
\affiliation{Inter-University Centre for Astronomy and Astrophysics, Pune 411007, India}
\author{F.~Acernese}
\affiliation{Universit\`a di Salerno, Fisciano, I-84084 Salerno, Italy}
\affiliation{INFN, Sezione di Napoli, Complesso Universitario di Monte S.Angelo, I-80126 Napoli, Italy}
\author{K.~Ackley}
\affiliation{OzGrav, School of Physics \& Astronomy, Monash University, Clayton 3800, Victoria, Australia}
\author{C.~Adams}
\affiliation{LIGO Livingston Observatory, Livingston, LA 70754, USA}
\author{V.~B.~Adya}
\affiliation{Max Planck Institute for Gravitational Physics (Albert Einstein Institute), D-30167 Hannover, Germany}
\affiliation{Leibniz Universit\"at Hannover, D-30167 Hannover, Germany}
\author{C.~Affeldt}
\affiliation{Max Planck Institute for Gravitational Physics (Albert Einstein Institute), D-30167 Hannover, Germany}
\affiliation{Leibniz Universit\"at Hannover, D-30167 Hannover, Germany}
\author{M.~Agathos}
\affiliation{University of Cambridge, Cambridge CB2 1TN, United Kingdom}
% ...

\collaboration{The LIGO Scientific Collaboration and the Virgo Collaboration}
  \maketitle
}

\clearpage
\appendix
\onecolumngrid
\newpage

\begin{center}
{\large\textbf{Supplement To: A search for the isotropic stochastic background in Advanced LIGO's second observing run}} \\[0.5cm]
(The LIGO Scientific Collaboration and Virgo Collaboration)
\end{center}
\vspace{0.55cm}
\twocolumngrid
\setcounter{equation}{0}
\renewcommand{\theequation}{A\arabic{equation}}

\section{Stochastic Upper Limit Construction}

Here we describe the parameter estimation procedure used to set upper limits on the amplitude of the stochastic background.

We assume a Gaussian likelihood for the measured cross-correlations $C(f)$ (see Eq.~\eqref{eq:def_cross_corr}), such that
\begin{equation}
\mathcal{L}(C | \Theta, \mathcal{H})
	\propto \exp\left(
   	-\frac{1}{2}\sum_k \frac{\left[C(f_k)-\langle C(\Theta; f_k)\rangle_\mathcal{H}\right]^2}{\sigma^2(f_k)}
       \right).
\end{equation}
Here, $\langle C(\Theta; f_k)\rangle_\mathcal{H}$ is the expected cross-correlation spectrum given hypothesis $\mathcal{H}$ and model parameters $\Theta$.
The variance $\sigma^2(f)$ is defined in Eq.~\eqref{eq:def_sigma}.
To perform parameter estimation using the combined results from Advanced LIGO's O1 and O2 observing runs, we use a joint likelihood of the form
\begin{equation}
\mathcal{L}(C_\mathrm{O1},C_\mathrm{O2} | \Theta, \mathcal{H}) = \mathcal{L}(C_\mathrm{O1} | \Theta, \mathcal{H}) \mathcal{L}(C_\mathrm{O2} | \Theta, \mathcal{H}).
\end{equation}

As discussed above, we adopt a power-law model for the energy-density spectrum of the stochastic background, giving the model spectrum $\langle C (f) \rangle = \Omega_\mathrm{ref}\left(f/f_\mathrm{ref}\right)^\alpha$.
We consider different two prior probability density distributions on $\Omega_\mathrm{ref}$: a uniform distribution $p(\Omega_{\rm ref}) \propto 1$ and a log-uniform distribution $p(\Omega_{\rm ref})\propto \Omega_{\rm ref}^{-1}$.
We use a peaked prior on the spectral index: $p(\alpha) \propto 1-\alpha/|\alpha_\mathrm{max}|$ with $\alpha_\mathrm{max} = 8$, preferring shallow slopes while still allowing for spectral indices far steeper than those predicted for known sources.

When allowing for alternative gravitational-wave polarizations, we will treat the total canonical energy density as a sum of three distinct power laws for the tensor, vector, and scalar-polarized components of the stochastic background.
In this case, our model cross-correlation spectrum becomes (see Eq.~\eqref{eq:nongrY})
\begin{equation}
\begin{aligned}
\langle C(f)\rangle_\mathrm{TVS} = \Omega^T_\mathrm{ref}(&f/f_\mathrm{ref})^{\alpha_T} + \beta_V(f)\Omega^V_\mathrm{ref}(f/f_\mathrm{ref})^{\alpha_V} \\
&+ \beta_S(f)\Omega^S_\mathrm{ref}(f/f_\mathrm{ref})^{\alpha_S},
\end{aligned}
\end{equation}
where $\Omega^A_\mathrm{ref}$ and $\alpha_A$ are the amplitudes and spectral indices of each polarization component. Note that $\beta_T(f)=\gamma_T(f)/\gamma_T(f)=1$.
Table \ref{tab:nongrUL} lists the 95\% credible upper limits on the amplitudes $\Omega^A_\mathrm{ref}$, following marginalization over spectral indices.

\section{Detailed Parameter Estimation Results}

\subsection{A. Standard Isotropic Search}

\begin{figure*}
\includegraphics[width=0.49\textwidth]{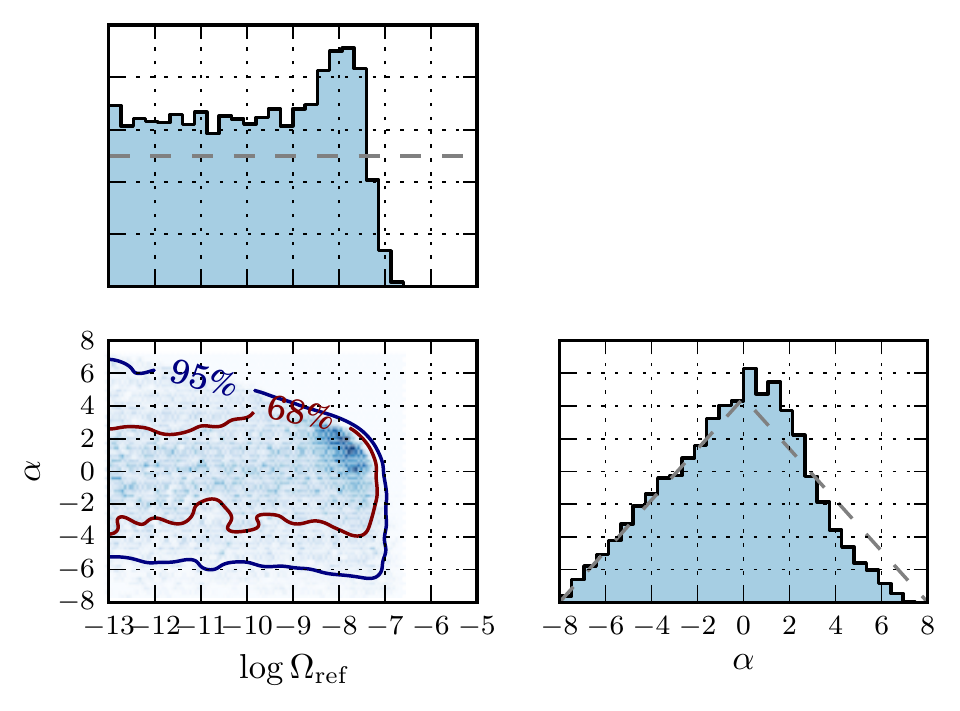}
\includegraphics[width=0.49\textwidth]{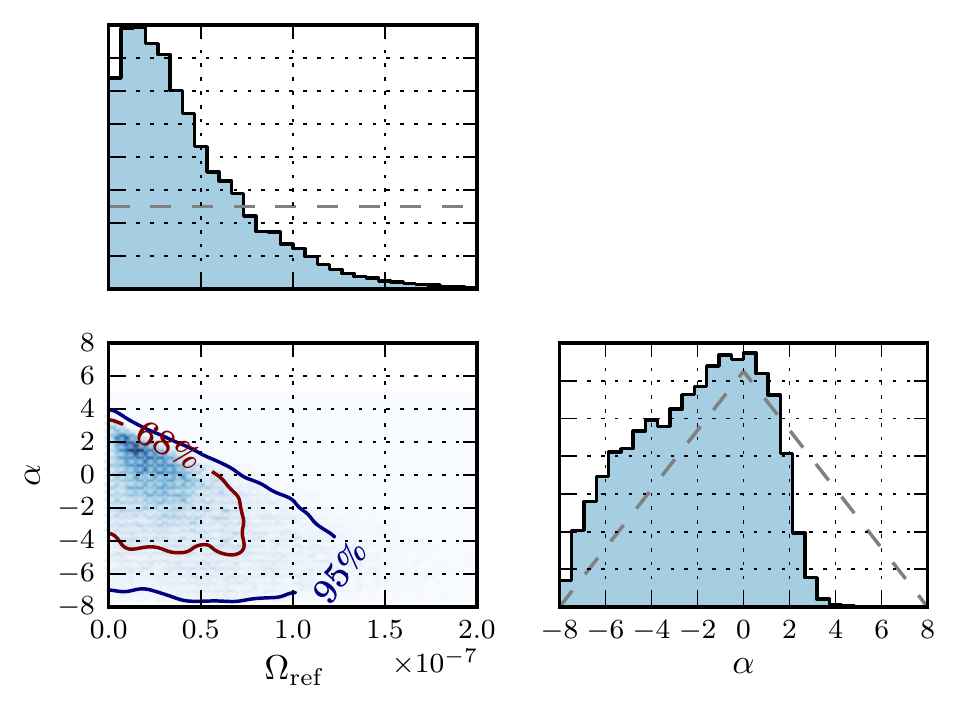}
\caption{Joint posteriors on the amplitude and spectral index of the isotropic stochastic background. The left plot assumes a log-uniform prior on the background's amplitude, while the right plot adopts a uniform amplitude prior. The dashed grey curves in the 1D marginalized posteriors mark the prior placed on each parameter, and the 2D posterior plots show contours containing the central 68\% and 95\% credible regions.
Marginalized 95\% credible bounds on $\Omega_\mathrm{ref}$ and $\alpha$ are listed in Table \ref{tab:isotropicFullPE}.
The $\Omega_\mathrm{ref}$ prior in the right plot has been multiplied by a factor of 50 so as to be visible.}
\label{fig:fullPosterior_T}
\end{figure*}

Figure \ref{fig:fullPosterior_T} shows the full posteriors obtained for the isotropic stochastic background's amplitude $\Omega_\mathrm{ref}$ and spectral index $\alpha$ using both O1 and O2 data, assuming both log-uniform (left) and uniform (right) amplitude priors.
Although the posteriors show a marginal feature at $\Omega_\mathrm{ref}\sim10^{-8}$ and $\alpha\sim 2$, this feature is not statistically significant, with a log Bayes factor of $\ln\mathcal{B}=\lnBayesTvsN$ between (tensor-polarized) signal and noise hypotheses.
The two choices of prior imply different constraints on $\alpha$, with the log-uniform prior yielding a symmetric spectral index posterior while the uniform prior results in a posterior skewed towards negative spectral indices.
For each choice of prior, Table \ref{tab:isotropicFullPE} lists the 95\% credible upper limits on $\Omega_\mathrm{ref}$, as well as the median and 95\% credible bounds on $\alpha$.

\begin{table}
\setlength{\tabcolsep}{18pt}
\renewcommand{\arraystretch}{1.3}
\begin{tabular}{l|c c}
\hline
\hline
Prior & $\Omega^{95\%}_\mathrm{ref}$ & $\alpha$ \\
\hline
Uniform
	& $\ngrRunOneTwoPolTUniformPriorLimitT$ 
    & $\ngrRunOneTwoPolTUniformPriorAlphaT
    	_{-\ngrRunOneTwoPolTUniformPriorAlphaTLowError}
        ^{+\ngrRunOneTwoPolTUniformPriorAlphaTHighError}$ \\
Log-uniform
	& $\ngrRunOneTwoPolTLogUniformPriorLimitT$
    & $\ngrRunOneTwoPolTLogUniformPriorAlphaT
    	_{-\ngrRunOneTwoPolTLogUniformPriorAlphaTLowError}
        ^{+\ngrRunOneTwoPolTLogUniformPriorAlphaTHighError}$ \\
\hline
\hline
\end{tabular}
\caption{95\% credible upper limits on $\Omega_\mathrm{ref}$ and the median recovered spectral index (with 95\% credible bounds) for both choices of amplitude prior, using combined data from Advanced LIGO's O1 and O2 observing runs.}
\label{tab:isotropicFullPE}
\end{table}

\subsection{B. Alternative Polarizations}

To compute odds $\mathcal{O}^\textsc{s}_\textsc{n}$ and $\mathcal{O}^\textsc{ngr}_\textsc{gr}$, we  independently consider each possible combination of gravitational-wave polarizations \cite{TestingGR_stoch,stoch_nongr_O1}.
Signal sub-hypothesis T, for example, allows only for tensor-polarized signals, while sub-hypothesis TV allows simultaneously for tensor and vector modes, etc.
In total, we must consider seven such sub-hypotheses: \{T,V,S,TV,TS,VS,TVS\}.
The union of all seven possibilities gives our signal hypothesis, while the union of the six hypotheses allowing vector or scalar modes gives our non-GR (NGR) hypothesis.
Table \ref{tab:modelEvidences} lists the log-Bayes factors found between each sub-hypothesis and Gaussian noise, using data from Advanced LIGO's O1 and O2 observing runs.

We must also choose the prior probabilities assigned to our various hypotheses.
We assign equal prior probabilities to the signal and noise hypotheses and weight each signal sub-hypothesis equally, giving
$\mathcal{O}^S_N = (1/7)\sum_\mathcal{H} \mathcal{B}^\mathcal{H}_\textsc{n}$ for $\mathcal{H}\in\mathrm{\{T,V,S,TV,TS,VS,TVS\}}$.
To obtain $\mathcal{O}^\textsc{ngr}_\textsc{gr}$ we similarly choose equal prior probabilities for the NGR and GR hypotheses and equal weights for each sub-hypothesis: $\mathcal{O}^\textsc{ngr}_\textsc{gr} = (1/6)\sum_\mathcal{H} \mathcal{B}^\mathcal{H}_\textsc{t}$ for $\mathcal{H}\in\mathrm{\{V,S,TV,TS,VS,TVS\}}$.

\vspace{0.1cm}
\begin{table}
\setlength{\tabcolsep}{12pt}
\renewcommand{\arraystretch}{1.3}
\begin{tabular}{l|r}
\hline
\hline
Hypothesis & $\ln\mathcal{B}$ \\
\hline
T & $\lnBayesTvsN$ \\
V & $\lnBayesVvsN$ \\
S & $\lnBayesSvsN$ \\
TV & $\lnBayesTVvsN$ \\
TS & $\lnBayesTSvsN$ \\
VS & $\lnBayesVSvsN$ \\
TVS & $\lnBayesTVSvsN$ \\
\hline
\hline
\end{tabular}
\caption{Log Bayes factors between each signal sub-hypothesis considered and the Gaussian noise hypothesis.}
\label{tab:modelEvidences}
\end{table}

\begin{figure*}
\includegraphics[width=0.95\textwidth]{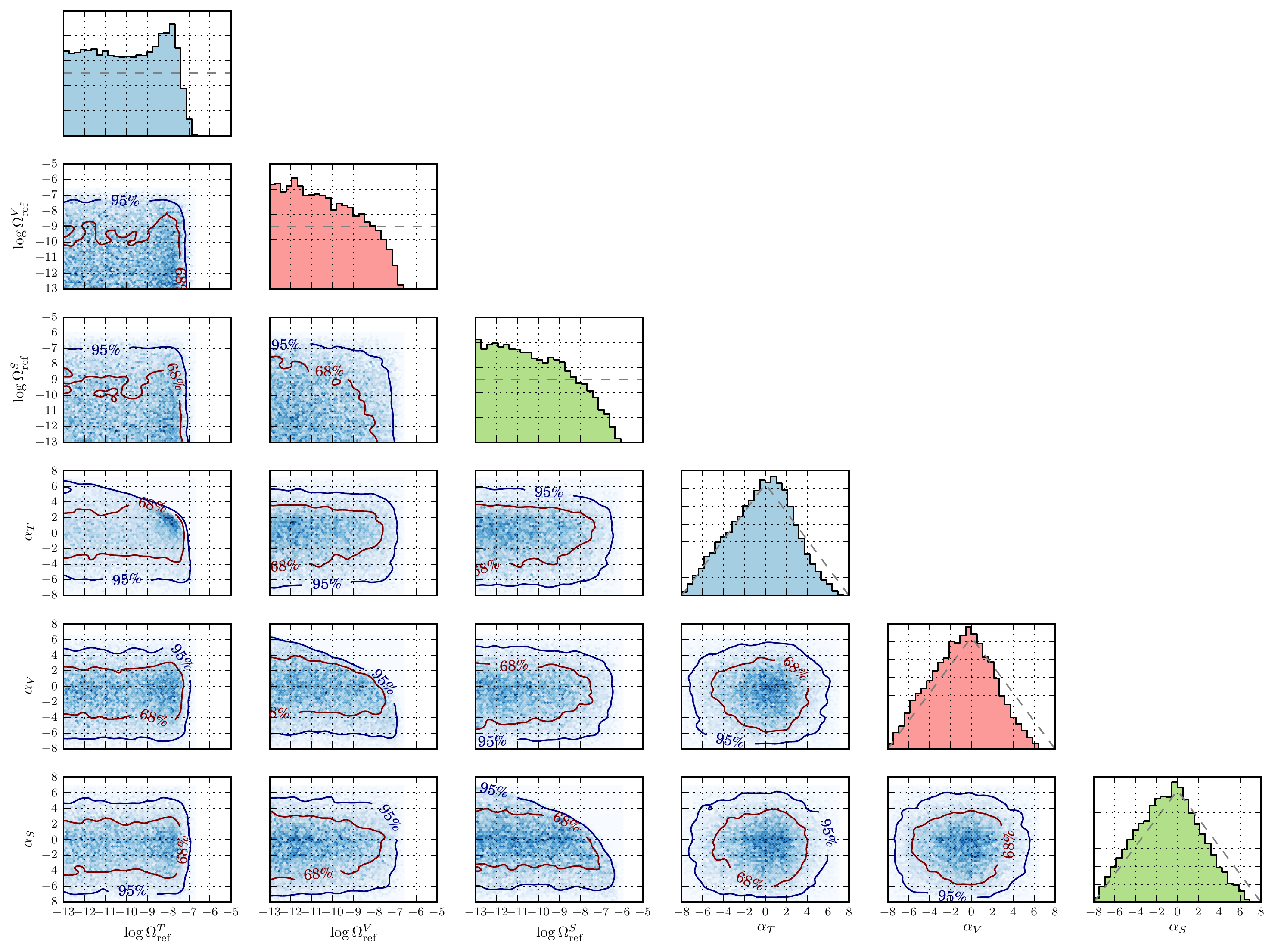}
\caption{Posteriors on the amplitudes and spectral indices of the tensor, vector, and scalar-polarized components of the stochastic background under the ``TVS'' hypothesis, assuming log-uniform amplitude priors.
Despite the marginal feature at $\log\Omega^T_\mathrm{ref}\sim-8$, these posteriors are consistent with Gaussian noise.
The dashed lines in each one-dimensional histogram mark the priors used, and each two-dimensional posterior shows contours encompassing the central 68\% and 95\% credible regions.
Marginalized 95\% credible upper limits on the amplitude parameters appear in Table \ref{tab:nongrUL}, as well as Table \ref{tab:ngrPEloguniform} below.
Table \ref{tab:ngrPEloguniform} also lists 95\% credible bounds on the three spectral indices.}
\label{fig:fullLogPosterior_TVS}
\end{figure*}

\begin{figure*}
\includegraphics[width=0.95\textwidth]{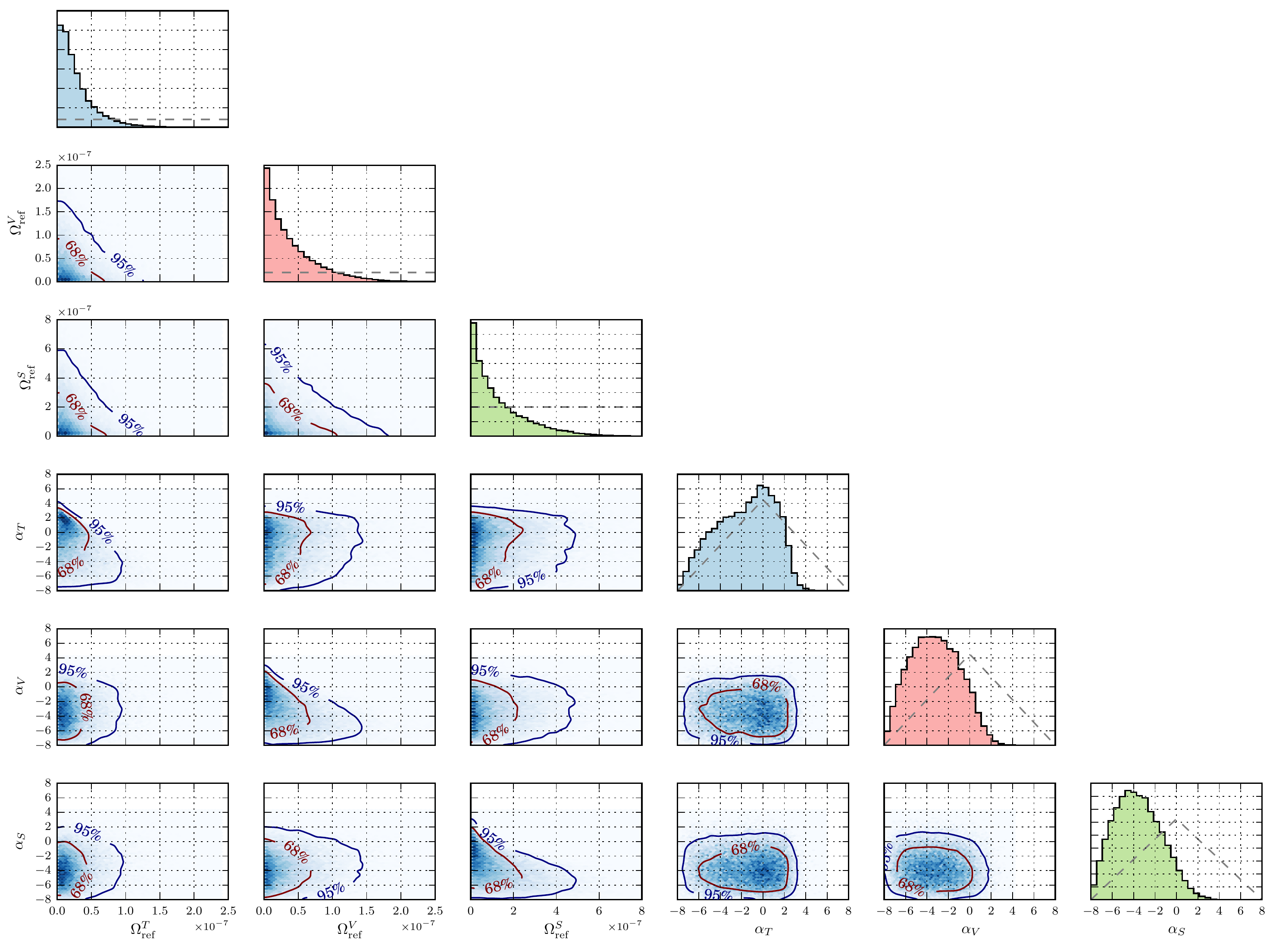}
\caption{As in Fig.~\ref{fig:fullLogPosterior_TVS}, but assuming uniform priors on the tensor, vector, and scalar stochastic background amplitudes.
The amplitude priors shown have been multiplied by a factor of 20 in order to be visible.
Marginalized parameter estimation results are listed below in Table \ref{tab:ngrPEuniform}.}
\label{fig:fullUniformPosterior_TVS}
\end{figure*}

In Table \ref{tab:nongrUL} above, we listed marginalized upper limits on possible tensor, vector, and scalar background amplitudes when allowing for the presence of all three polarization types (the TVS hypothesis).
Figures \ref{fig:fullLogPosterior_TVS} and \ref{fig:fullUniformPosterior_TVS} show the corresponding six-dimensional posteriors on all amplitudes and spectral indices, under both log-uniform and uniform amplitude priors.
The slight amplitude peak discussed above is also visible in Fig.~\ref{fig:fullLogPosterior_TVS}.
Interestingly, while one might expect a loud noise realization to contribute indiscriminately to all three polarization sectors, only the $\Omega^T_\mathrm{ref}$ posterior is peaked.
Once again, though, this peak is not statistically significant, and the posteriors remain consistent with Gaussian noise.

Tables \ref{tab:ngrPEloguniform} and \ref{tab:ngrPEuniform} list marginalized limits on the amplitudes and spectral indices of tensor, vector, and scalar-polarized stochastic backgrounds for each signal hypothesis allowing alternative polarizations.

\begin{table*}
\setlength{\tabcolsep}{18pt}
\renewcommand{\arraystretch}{1.3}
\begin{tabular}{l|c c c c c c}
\hline
\hline
Hypothesis & $\Omega^{T,95\%}_\mathrm{ref}$ & $\Omega^{V,95\%}_\mathrm{ref}$ & $\Omega^{S,95\%}_\mathrm{ref}$
	& $\alpha_T$ & $\alpha_V$ & $\alpha_S$ \\
\hline
V 
	& - 
    & $\ngrRunOneTwoPolVLogUniformPriorLimitV$ 
    & - 
	& - 
    & $\ngrRunOneTwoPolVLogUniformPriorAlphaV
    	_{-\ngrRunOneTwoPolVLogUniformPriorAlphaVLowError}
        ^{+\ngrRunOneTwoPolVLogUniformPriorAlphaVHighError}$
    & - \\
S 
	& - 
    & -
    & $\ngrRunOneTwoPolSLogUniformPriorLimitS$ 
	& - 
    & -
    & $\ngrRunOneTwoPolSLogUniformPriorAlphaS
    	_{-\ngrRunOneTwoPolSLogUniformPriorAlphaSLowError}
        ^{+\ngrRunOneTwoPolSLogUniformPriorAlphaSHighError}$ \\
TV 
	& $\ngrRunOneTwoPolTVLogUniformPriorLimitT$ 
    & $\ngrRunOneTwoPolTVLogUniformPriorLimitV$ 
    & - 
    & $\ngrRunOneTwoPolTVLogUniformPriorAlphaT
    	_{-\ngrRunOneTwoPolTVLogUniformPriorAlphaTLowError}
        ^{+\ngrRunOneTwoPolTVLogUniformPriorAlphaTHighError}$
    & $\ngrRunOneTwoPolTVLogUniformPriorAlphaV
    	_{-\ngrRunOneTwoPolTVLogUniformPriorAlphaVLowError}
        ^{+\ngrRunOneTwoPolTVLogUniformPriorAlphaVHighError}$
    & - \\
TS 
	& $\ngrRunOneTwoPolTSLogUniformPriorLimitT$ 
    & -
    & $\ngrRunOneTwoPolTSLogUniformPriorLimitS$  
    & $\ngrRunOneTwoPolTSLogUniformPriorAlphaT
    	_{-\ngrRunOneTwoPolTSLogUniformPriorAlphaTLowError}
        ^{+\ngrRunOneTwoPolTSLogUniformPriorAlphaTHighError}$
    & -
    & $\ngrRunOneTwoPolTSLogUniformPriorAlphaS
    	_{-\ngrRunOneTwoPolTSLogUniformPriorAlphaSLowError}
        ^{+\ngrRunOneTwoPolTSLogUniformPriorAlphaSHighError}$ \\
VS 
	& -
    & $\ngrRunOneTwoPolVSLogUniformPriorLimitV$
    & $\ngrRunOneTwoPolVSLogUniformPriorLimitS$  
    & -
    & $\ngrRunOneTwoPolVSLogUniformPriorAlphaV
    	_{-\ngrRunOneTwoPolVSLogUniformPriorAlphaVLowError}
        ^{+\ngrRunOneTwoPolVSLogUniformPriorAlphaVHighError}$
    & $\ngrRunOneTwoPolVSLogUniformPriorAlphaS
    	_{-\ngrRunOneTwoPolVSLogUniformPriorAlphaSLowError}
        ^{+\ngrRunOneTwoPolVSLogUniformPriorAlphaSHighError}$ \\
TVS 
	& $\ngrRunOneTwoPolTVSLogUniformPriorLimitT$
    & $\ngrRunOneTwoPolTVSLogUniformPriorLimitV$
    & $\ngrRunOneTwoPolTVSLogUniformPriorLimitS$  
    & $\ngrRunOneTwoPolTVSLogUniformPriorAlphaT
    	_{-\ngrRunOneTwoPolTVSLogUniformPriorAlphaTLowError}
        ^{+\ngrRunOneTwoPolTVSLogUniformPriorAlphaTHighError}$
    & $\ngrRunOneTwoPolTVSLogUniformPriorAlphaV
    	_{-\ngrRunOneTwoPolTVSLogUniformPriorAlphaVLowError}
        ^{+\ngrRunOneTwoPolTVSLogUniformPriorAlphaVHighError}$
    & $\ngrRunOneTwoPolTVSLogUniformPriorAlphaS
    	_{-\ngrRunOneTwoPolTVSLogUniformPriorAlphaSLowError}
        ^{+\ngrRunOneTwoPolTVSLogUniformPriorAlphaSHighError}$ \\
\hline
\hline
\end{tabular}
\caption{Marginalized parameter estimation results obtained assuming each unique combination of tensor, vector, and scalar polarizations, assuming log-uniform amplitude priors (the tensor-only case appears above in Table \ref{tab:isotropicFullPE}.
Specifically, we list the 95\% credible upper limits on each amplitude parameter, and the median recovered spectral indices with uncertainties encompassing the central 95\% credible region.}
\label{tab:ngrPEloguniform}
\end{table*}

\begin{table*}
\setlength{\tabcolsep}{18pt}
\renewcommand{\arraystretch}{1.3}
\begin{tabular}{l|c c c c c c}
\hline
\hline
Hypothesis & $\Omega^{T,95\%}_\mathrm{ref}$ & $\Omega^{V,95\%}_\mathrm{ref}$ & $\Omega^{S,95\%}_\mathrm{ref}$
	& $\alpha_T$ & $\alpha_V$ & $\alpha_S$ \\
\hline
V 
	& - 
    & $\ngrRunOneTwoPolVUniformPriorLimitV$ 
    & - 
	& - 
    & $\ngrRunOneTwoPolVUniformPriorAlphaV
    	_{-\ngrRunOneTwoPolVUniformPriorAlphaVLowError}
        ^{+\ngrRunOneTwoPolVUniformPriorAlphaVHighError}$
    & - \\
S 
	& - 
    & -
    & $\ngrRunOneTwoPolSUniformPriorLimitS$ 
	& - 
    & -
    & $\ngrRunOneTwoPolSUniformPriorAlphaS
    	_{-\ngrRunOneTwoPolSUniformPriorAlphaSLowError}
        ^{+\ngrRunOneTwoPolSUniformPriorAlphaSHighError}$ \\
TV 
	& $\ngrRunOneTwoPolTVUniformPriorLimitT$ 
    & $\ngrRunOneTwoPolTVUniformPriorLimitV$ 
    & - 
    & $\ngrRunOneTwoPolTVUniformPriorAlphaT
    	_{-\ngrRunOneTwoPolTVUniformPriorAlphaTLowError}
        ^{+\ngrRunOneTwoPolTVUniformPriorAlphaTHighError}$
    & $\ngrRunOneTwoPolTVUniformPriorAlphaV
    	_{-\ngrRunOneTwoPolTVUniformPriorAlphaVLowError}
        ^{+\ngrRunOneTwoPolTVUniformPriorAlphaVHighError}$
    & - \\
TS 
	& $\ngrRunOneTwoPolTSUniformPriorLimitT$ 
    & -
    & $\ngrRunOneTwoPolTSUniformPriorLimitS$  
    & $\ngrRunOneTwoPolTSUniformPriorAlphaT
    	_{-\ngrRunOneTwoPolTSUniformPriorAlphaTLowError}
        ^{+\ngrRunOneTwoPolTSUniformPriorAlphaTHighError}$
    & -
    & $\ngrRunOneTwoPolTSUniformPriorAlphaS
    	_{-\ngrRunOneTwoPolTSUniformPriorAlphaSLowError}
        ^{+\ngrRunOneTwoPolTSUniformPriorAlphaSHighError}$ \\
VS 
	& -
    & $\ngrRunOneTwoPolVSUniformPriorLimitV$
    & $\ngrRunOneTwoPolVSUniformPriorLimitS$  
    & -
    & $\ngrRunOneTwoPolVSUniformPriorAlphaV
    	_{-\ngrRunOneTwoPolVSUniformPriorAlphaVLowError}
        ^{+\ngrRunOneTwoPolVSUniformPriorAlphaVHighError}$
    & $\ngrRunOneTwoPolVSUniformPriorAlphaS
    	_{-\ngrRunOneTwoPolVSUniformPriorAlphaSLowError}
        ^{+\ngrRunOneTwoPolVSUniformPriorAlphaSHighError}$ \\
TVS 
	& $\ngrRunOneTwoPolTVSUniformPriorLimitT$
    & $\ngrRunOneTwoPolTVSUniformPriorLimitV$
    & $\ngrRunOneTwoPolTVSUniformPriorLimitS$  
    & $\ngrRunOneTwoPolTVSUniformPriorAlphaT
    	_{-\ngrRunOneTwoPolTVSUniformPriorAlphaTLowError}
        ^{+\ngrRunOneTwoPolTVSUniformPriorAlphaTHighError}$
    & $\ngrRunOneTwoPolTVSUniformPriorAlphaV
    	_{-\ngrRunOneTwoPolTVSUniformPriorAlphaVLowError}
        ^{+\ngrRunOneTwoPolTVSUniformPriorAlphaVHighError}$
    & $\ngrRunOneTwoPolTVSUniformPriorAlphaS
    	_{-\ngrRunOneTwoPolTVSUniformPriorAlphaSLowError}
        ^{+\ngrRunOneTwoPolTVSUniformPriorAlphaSHighError}$ \\
\hline
\hline
\end{tabular}
\caption{As in Table \ref{tab:ngrPEloguniform}, but assuming uniform amplitude priors.}
\label{tab:ngrPEuniform}
\end{table*}

\subsection{Low noise LEMI magnetometers}
\begin{figure}
 \includegraphics[width=0.49\textwidth]{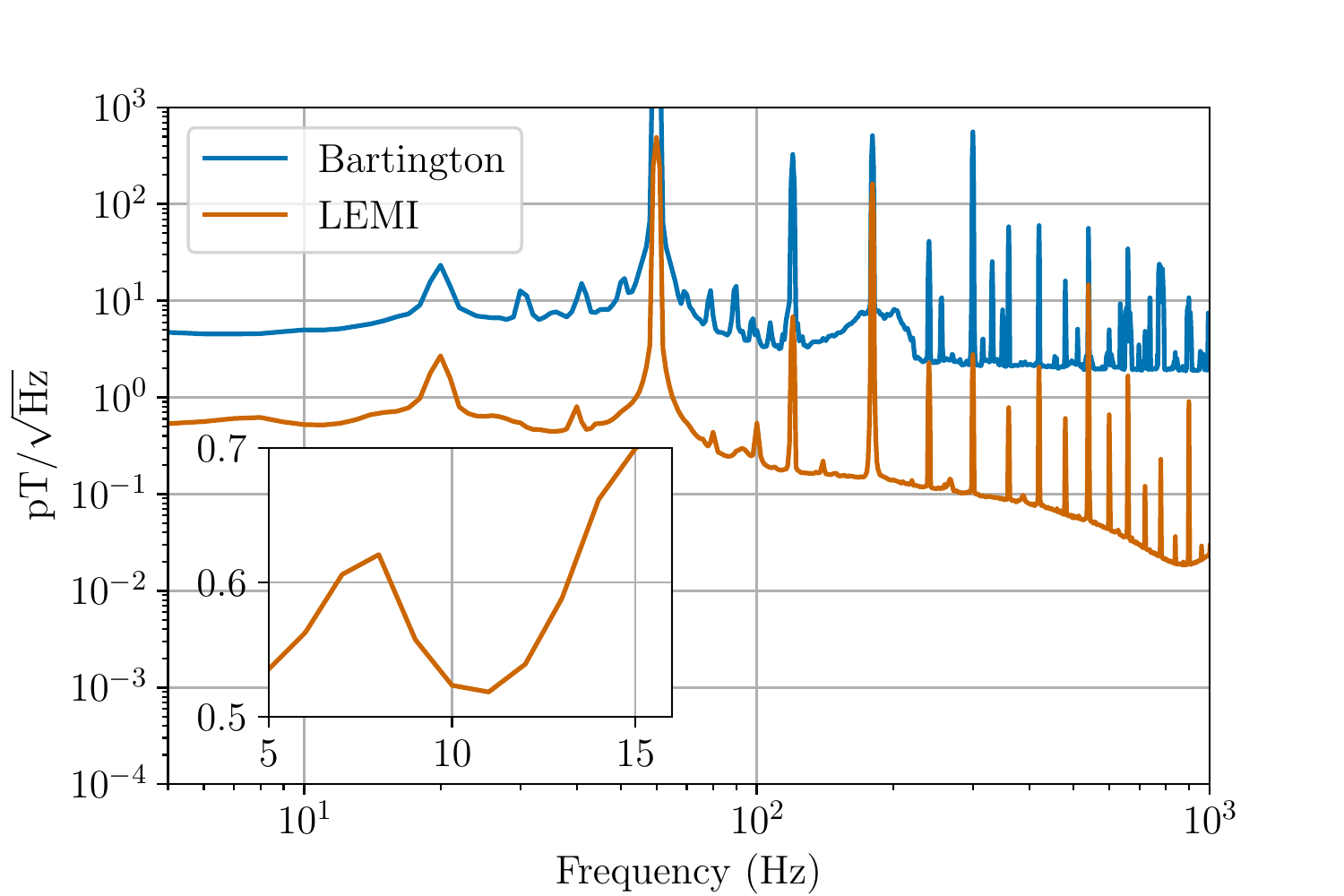}
 \caption{Power spectral densities for the Bartington and LEMI magnetometers. We choose one hour of data, March 4 2017 06:00-07:00, which was a time with low magnetic noise, to compare the instruments. The inset shows the LEMI spectrum from 5-15 Hz, where the first Schumann resonance  at $\sim$ 8 Hz is clearly visible.}
 \label{fig:bart_vs_lemi}
 \end{figure}

In future observing runs, correlated magnetic noise may require subtraction techniques \cite{Schumann_1,Schumann_2,Schumann_3,Schumann_4} or parameter estimation \cite{meyers_thesis}. Sensitive LEMI-120 magnetometers \cite{lemi} have been installed at both sites. In Figure \ref{fig:bart_vs_lemi}, we show the power spectral density for a Bartington and a LEMI magnetometer at the Livingston site, using one hour of data during a day with low magnetic noise. The lower noise floor of the LEMI magnetometer can be used to perform a better measurement of the Schumann resonances in real time. We emphasize that this figure is only meant to compare the instruments. An accurate estimate of magnetic noise  must compute the coherence over the entire run; the results of this analysis using Bartington magnetometers is shown in Figure \ref{fig:schumann}.
 
\end{document}